\documentclass[apj]{emulateapj}

\usepackage{graphicx}
\usepackage{newtxtext,newtxmath}
\usepackage{psfig}
\usepackage{natbib}
\usepackage{color}
\usepackage{soul}

\def\asec{\ifmmode ^{\prime\prime}\else$^{\prime\prime}$\fi}

\def\msun{\hbox{M$_{\odot}$}}

\def\msunyr{\mbox{\,${\rm M_{\odot}\, yr^{-1}}$}}
\def\mdot{\dot M}

\def\degs{\ifmmode ^{\circ}\else$^{\circ}$\fi}
\def\amin{\ifmmode ^{\prime}\else$^{\prime}$\fi}
\def\asec{\ifmmode ^{\prime\prime}\else$^{\prime\prime}$\fi}
\def\degs{\ifmmode ^{\circ}\else$^{\circ}$\fi}
\def\amin{\ifmmode ^{\prime}\else$^{\prime}$\fi}

\def\EE#1{\times 10^{#1}}

\def\cm{\mbox{\,cm}}

\def\cm3{\mbox{\,cm$^{-3}$}}
\def\kms{\mbox{\,km~s$^{-1}$}}

\def\kms{\mbox{\,km s$^{-1}$}}

\def\lsim{\!\!\!\phantom{\le}\smash{\buildrel{}\over
 {\lower2.5dd\hbox{$\buildrel{\lower2dd\hbox{$\displaystyle<$}}\over
                                 \sim$}}}\,\,}
\def\gsim{\!\!\!\phantom{\ge}\smash{\buildrel{}\over
{\lower2.5dd\hbox{$\buildrel{\lower2dd\hbox{$\displaystyle>$}}\over
                               \sim$}}}\,\,}
\def\albanova{1}
\def\oskar{2}
\def\iaa{4}
\def\unizar{5}
\def\kvali{6}
\def\itep{7}
\def\vniia{8}
\def\sai{3}
\def\ncra{9}

\shorttitle{Evolution of the progenitors of SNe 1993J and 2011dh revealed through late time radio and X-ray studies}
\shortauthors{Kundu et al.}

\begin{document}

\title{Evolution of the progenitors of SNe 1993J and 2011dh revealed through late time radio and X-ray studies}

\author{E. Kundu \altaffilmark{\albanova,\oskar},
P. Lundqvist \altaffilmark{\albanova,\oskar},  
%C-I. Bj\"ornsson \altaffilmark{\albanova,\oskar}, 
E. Sorokina \altaffilmark{\sai},
M.~A. P\'erez-Torres  \altaffilmark{\iaa,\unizar},  
S. Blinnikov \altaffilmark{\kvali,\itep,\vniia},
%E. Sorokina \altaffilmark{\sai},
 E. O'Connor \altaffilmark{\albanova,\oskar},
 M. Ergon \altaffilmark{\albanova,\oskar},
 P. Chandra \altaffilmark{\albanova,\ncra},
 B. Das \altaffilmark{\ncra}
%R. Herrero-Illana \altaffilmark{\eso}, 
%A. Alberdi \altaffilmark{\iaa}
}

%\altaffiltext{\albanova}{Department of Astronomy and The Oskar Klein Centre, AlbaNova, Stockholm University, SE-10691 Stockholm, Sweden.\\
%esha.kundu@astro.su.se
%}
\altaffiltext{\albanova}{Department of Astronomy AlbaNova, Stockholm University, SE-10691 Stockholm, Sweden.\\
esha.kundu@astro.su.se
}
\altaffiltext{\oskar}{The Oskar Klein Centre, AlbaNova, SE-10691 Stockholm, Sweden.}
\altaffiltext{\iaa}{Instituto de Astrof\'isica de Andaluc\'ia, Glorieta de las Astronom\'ia, s/n, E-18008 Granada, Spain.}
\altaffiltext{\unizar}{Visiting Scientist: Departamento de F\'isica Teorica, Facultad de Ciencias, Universidad de Zaragoza, Spain.}
\altaffiltext{\kvali}{Kavli Institute for the Physics and Mathematics of the Universe (WPI), University of Tokyo.}
\altaffiltext{\itep}{Institute for Theoretical and Experimental Physics (ITEP), Moscow, Russia.}
\altaffiltext{\vniia}{All-Russia Research Institute of Automatics (VNIIA), Moscow, Russia.}
\altaffiltext{\sai}{Sternberg Astronomical Institute, M.V. Lomonosov Moscow State University, Universitetski pr. 13, 119234 Moscow, Russia}
\altaffiltext{\ncra}{National Centre for Radio Astronomy, Tata Institute of Fundamental Research, Pune University Campus, Pune-411007, India}
%\altaffiltext{\eso}{European Southern Observatory (ESO), Alonso de C\'ordova 3107, Vitacura, Casilla 19001, Santiago de Chile, Chile.}
%\altaffiltext{\oskar}{The Oskar Klein Centre, AlbaNova, SE-10691 Stockholm, Sweden.}

\begin{abstract}

We perform hydrodynamical simulations of the interaction between supernova (SN) ejecta and circumstellar medium (CSM) for SN~1993J and SN~2011dh, and calculate the radio and X-ray emissions expected from the shocked gas at late epochs ($t$). Considering the ejecta structure from multi-group radiation hydrodynamics simulation, we find that the observed rapid drop in radio and X-ray light curves of SN~1993J at $t>$3000~days can be due to a change in the mass-loss rate ($\mdot$) around $\sim$6500 years prior to the explosion of the SN. The exact epoch scales inversely with the assumed wind velocity of $v_{\rm w}=10\kms$. The progenitor of this SN very likely belonged to a binary system, where, during its evolution, the primary had transferred material to the secondary. It is argued in the paper that the change in $\mdot$ can happen because of a change in the mass accretion efficiency ($\eta$) of the companion star. It is possible that before $\sim6500~(v_{\rm w}/10~\kms)^{-1}$~years prior to the explosion, $\eta$ was high, thus the CSM was tenuous, which causes the late time downturn in fluxes. In the case of SN~2011dh, the late time evolution is found to be consistent with a wind medium with $\mdot/v_w=4\times10^{-6}\msunyr/10\kms$. It is difficult from our analysis to predict whether the progenitor of this SN had a binary companion, however, if future observations show similar decrease in radio and X-ray fluxes, then that would give strong support to a scenario where both SNe had undergone similar kind of binary evolution before explosion.

\end{abstract}


\section{Introduction}
\label{sec:intro}
Supernovae (SNe) are massive destruction of stars. The evolution of these stars before explosion depends on the fact whether or not it had interaction with a companion star during its life time \citep{pod92, pod01}. In most of the cases it is not possible to verify, due to large astronomical distances, if the exploded star was part of a binary system. However, indirect evidence, such as signs of hydrogen emission at early time and He I absorption at late epochs, or no trace of hydrogen line in the optical spectra, point toward a progenitor which has been stripped of most of, or the entire envelope, because of an interaction with a secondary star \citep{nomoto93, pod93}. Nevertheless, massive stars, with a main sequence mass $> 20 \msun$, in isolation, can undergo huge mass loss, driven by strong winds \citep{sch92, vink05} and periodic eruptions \citep{smith06}, which can potentially remove the very outermost layer, partially or completely. But models suggest that the stars which end their lives with very low envelope mass fall within a very narrow initial mass range (see  \citet{smartt11} for a review on missing high mass progenitors). Furthermore, the stripped envelope SN rate is not consistent with the fraction of massive single-stars that are expected to be accounted for  this kind of SNe. It is therefore more probable that the majority of the stripped envelope progenitors are low mass stars and must originate from a binary system %have had a companion star
\citep{smith11}. In this case, during the evolution, the primary fills its Roche lobe and transfers material to the companion. It is found that Roche lobe overflow is an effective mechanism through which stars can lose most of their envelope mass \citep{pod92, pod01}. 
%Among the strip envelope SNe progenitors, the stars with 0.1 - 1.0 $\msun$ hydrogen envelope left at the time of explosion, display, in their post explosion optical spectra, the hydrogen lines at early epoch and after few weeks/months the He I absorption feature. Therefore, these explosions are categorized as SN type IIb (see the review by \citet{fill97}). Therefore, these explosions are categorized as SN type IIb (see the review by \citet{fill97}).
 Among the stripped envelope SNe, which display in their optical spectra hydrogen lines at early epoch and after few weeks/months the He I absorption feature, are categorized as SN type IIb (see the review by \citet{fill97}).

\par 
Beside SN 2008ax \citep{crock08} and SN 2013df \citep{van14} the other two SNe IIb, for which the progenitor stars have been directly identified by analyzing the pre-explosion archival images, are SN 1993J \citep{cohen95} and indeed SN 2011dh \citep{van11}. For SN 1993J the optical light curve acquired just after the explosion revealed the presence of two peaks, which were best explained by a binary model with both stars having a mass of $\sim$ 15 $\msun$ \citep{nomoto93, pod93}. This scenario is further confirmed by \citet{aldering94} by studying the observed photometry of the SN, and the authors concluded that the progenitor star was a G8-K0 supergiant. Furthermore, investigating the ultraviolet images taken using the Hubble Space telescope (HST), \citet{maund04} justified the presence of a hot B type companion star in the vicinity of the SN.  
%and propose \citet{maund09} that the companion would be visible by 2012 if the B and U band flux from the SN location decreases according to the current rate. Though the companion has so far                 
Later \citet{fox14} have drawn a similar conclusion from the observations carried out at the far ultraviolet spectral region. It is true that the companion of SN 1993J has so far avoided the discovery, but it is almost certain that the progenitor of this SN was part of a binary system. In case of SN 2011dh a yellow supergiant (YSG) star was detected at the location of the SN by examining the pre-explosion HST archival images \citep{van11}. Initially, this star was thought to be the binary companion of the SN or an unrelated star \citep{van11, arcavi11, sod12}, as the  putative progenitor was found to be less extended compared to the detected YSG. However, using  pre- and post-explosion HST and Gemini images, \citet{maund11,van13} established the fact that the YSG was the progenitor of SN 2011dh. The early spectra of SN 2011dh has shown the presence of a thin layer of envelope \citep{arcavi11a, marion11} at the time of the explosion. Therefore, it could be expected that the progenitor of this SN belonged to a binary system. However, it is not fully ruled out that it was instead an explosion of a star that had evolved in isolation \citep{georgy11}. %any excess in UV has been observed from the Like SN1993J           

\par
In this regard, the medium surrounding SNe 1993J and 2011dh provides important clues about the evolution of both SNe. In a binary system, the primary usually transfers mass to the secondary through Roche lobe overflow. According to the accretion efficiency some fraction of this mass gets lost into the ambient medium. After the explosion, the supersonic SN ejecta interact with this medium and shocks are formed. These shocks channel part of their kinetic energy to accelerate charged particles into relativistic particles.
%This discontinuities are ideal places where charged particles can potentially be accelerated to relativistic energies. 
As magnetic fields also get amplified in the shocks \citep{bykov13, cap14b, kundu17}, these accelerated particles, mainly electrons, lose a fraction of their energy through synchrotron emission. This radiation often makes the SN shocks bright in radio frequencies. The intensity of this brightness depends on the density of the CSM, which means that by studying this emission one could map the mass loss history of the progenitor system, and hence can gain information about the system. The X-ray emission, on the other hand, at late epoch mainly comes from the shocked ejecta behind the reverse shock. There are different mechanisms, i.g., free-free, free-bound, two-photons, line emissions, which contribute to the emission at late times and provide important clues about the circumbinary medium.       

\par 
The radio and X-ray observations of SN 1993J have revealed a sudden downturn in radio and X-ray fluxes beyond $\sim$ 3000 days. We study here the evolution of the radio and the X-ray fluxes, from this SN, by performing the hydrodynamical simulations of the interaction of the SN ejecta with CSM, and calculating the fluxes using a post-processing procedure. As SN 2011dh is similar to SN 1993J in many aspects, a similar analysis has been carried out for this SN. 

%Radio and X-ray observations of SN 1993J have revealed a sudden downturn in radio and X-ray fluxes beyond $\sim$ 3000 days. We studied the evolution of the radio and the X-ray fluxes by performing hydrodynamical simulations of the interaction of the SN ejecta with the CSM, and calculating the fluxes using a post-processing procedure. From our study it is found that before $t_{\rm chng} \sim 6500~(v_{\rm w} / 10~\kms)^{-1}$ years prior to the explosion, the mass loss rate of the progenitor system was very low, however around $t_{\rm chng}$ up to the explosion, the system has lost mass at a much higher rate. In terms of accretion efficiency ($\eta$) of the binary companion this suggests that much before $t_{\rm chng}$, $\eta$ was very high. Therefore, only a small percentage of the mass gets ejected in the ambient medium. However, around $t_{\rm chng}$ the accretion efficiency decreases (the exact reason of this decrease is not known). As a result a much denser CSM is created.       

\par 
%As SN 2011dh is similar to SN 1993J in many aspects, a similar analysis has been been carried out for this SN. It is found from our analysis that until now (up to the time we have observed data) the evolution of this SN is consistent with a wind medium. However our analysis points toward a denser medium, characterized by a mass loss rate of $4 \times 10^{-6}$ $\msunyr$ for a wind velocity of 10 $\kms$, compared to what has been obtained from the earlier radio \citep{sod12, krauss12} and X-ray \citep{maeda14} studies. 

Along with the CSM, the SN ejecta structure plays an important role in the evolution of the SN. Therefore, we  considered the ejecta structure of SN 1993J and SN 2011dh from numerical simulations, which use the multi-group radiation hydrodynamic simulations (STELLA) to evaluate the structures. The details about the STELLA simulations and ejecta structure are discussed in detail in $\S$ \ref{sec:ejecta_model}. The rest of the paper is arranged in the following way. In $\S$ \ref{sec:hydro_sim} we present the hydrodynamical simulations of the interaction between the SN ejecta and the CSM. Afterwards, $\S$ \ref{sec:data},  \ref{sec:radio_emission} and \ref{sec:Xray_emission} give an account of radio and X-ray data, modeling of radio and X-ray emission, respectively. Next the results are presented and discussed in $\S$ \ref{sec:results} and \ref{sec:discussion}, respectively. At last, the conclusions are drawn in $\S$ \ref{sec:conclusions}.

%the shocks convert some amount of its kinetic energy into the relativistic particles           

%a star carries important clues about evolution of that star. During the lifetime stars lose mass that gets ejected into the ambient medium. In many cases the mass loss is driven by winds, therefore the density of the circumstellar medium (CSM), $\rho_{\rm csm}$, decreases with $r^{-2}$, where $r$ is the radial distance. In binary system the          

%Although it is very rare but there are four nearby SNe type IIb, SN 1993J\citep{}, SN 2008ax \citep{}, SN 2011dh \citep{} and SN 2013dh \citep{} ,for which the progenitor stars have been directly identified by analyzing the pre-explosion archival images. The In this paper we focus on the evolution of the progenitors of SN 1993J and SN 2011dh through radio and X-ray analysis.  

%Among these four SN 1993J and SN 2011dh have been extensively observed in radio and X-ray wavelengths.    

%distance SN 2008ax 7.7 Mpc, Sn 2013df 16.6 Mpc

%It is usually assumed that before explosion these stars have evolved in isolation or could be part of a binary system. In binary system the e  

\section{Ejecta Model}
\label{sec:ejecta_model}
\begin{figure*}
\centering
\includegraphics[width=8cm,angle=0]{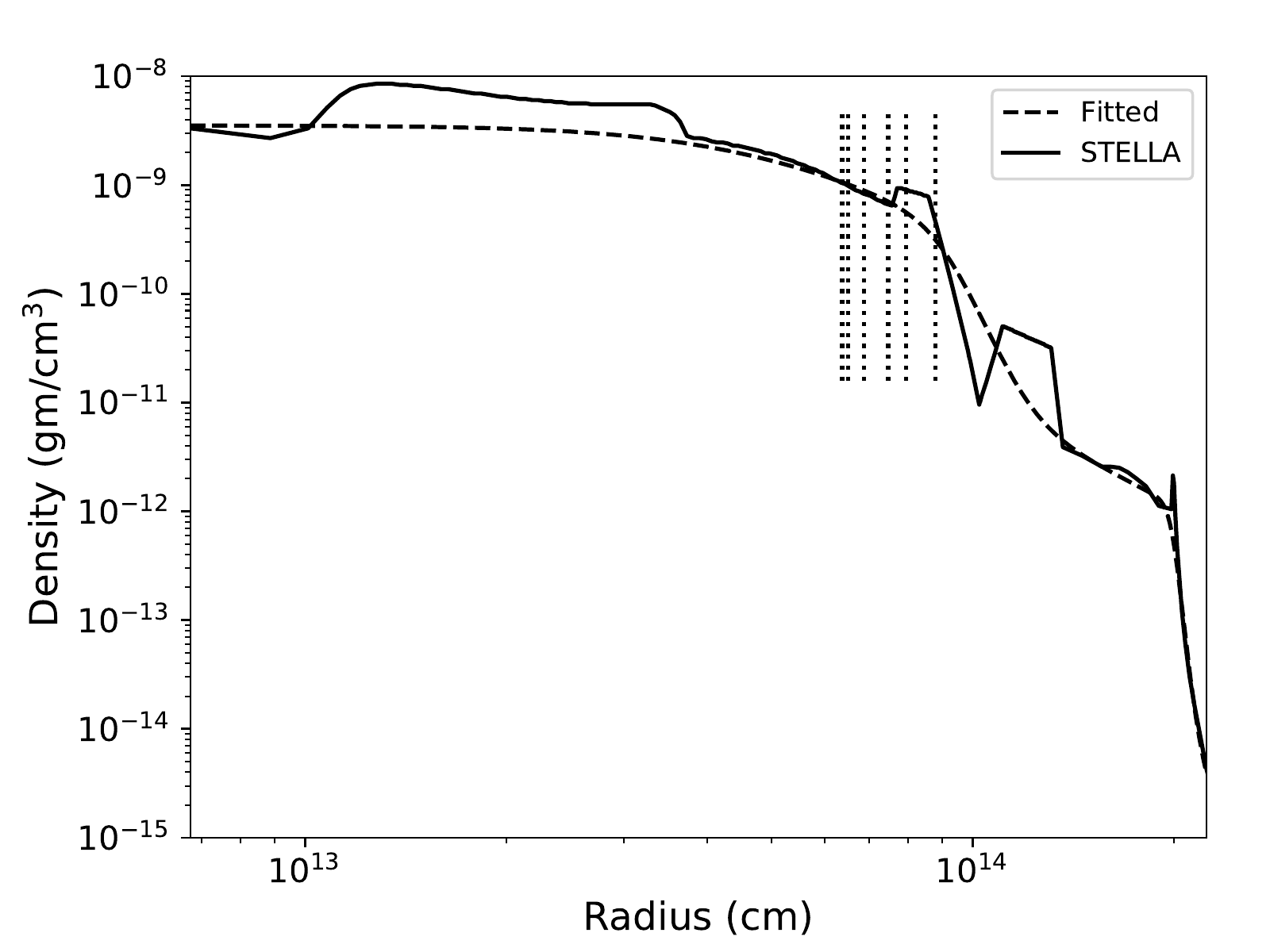}
\includegraphics[width=8cm,angle=0]{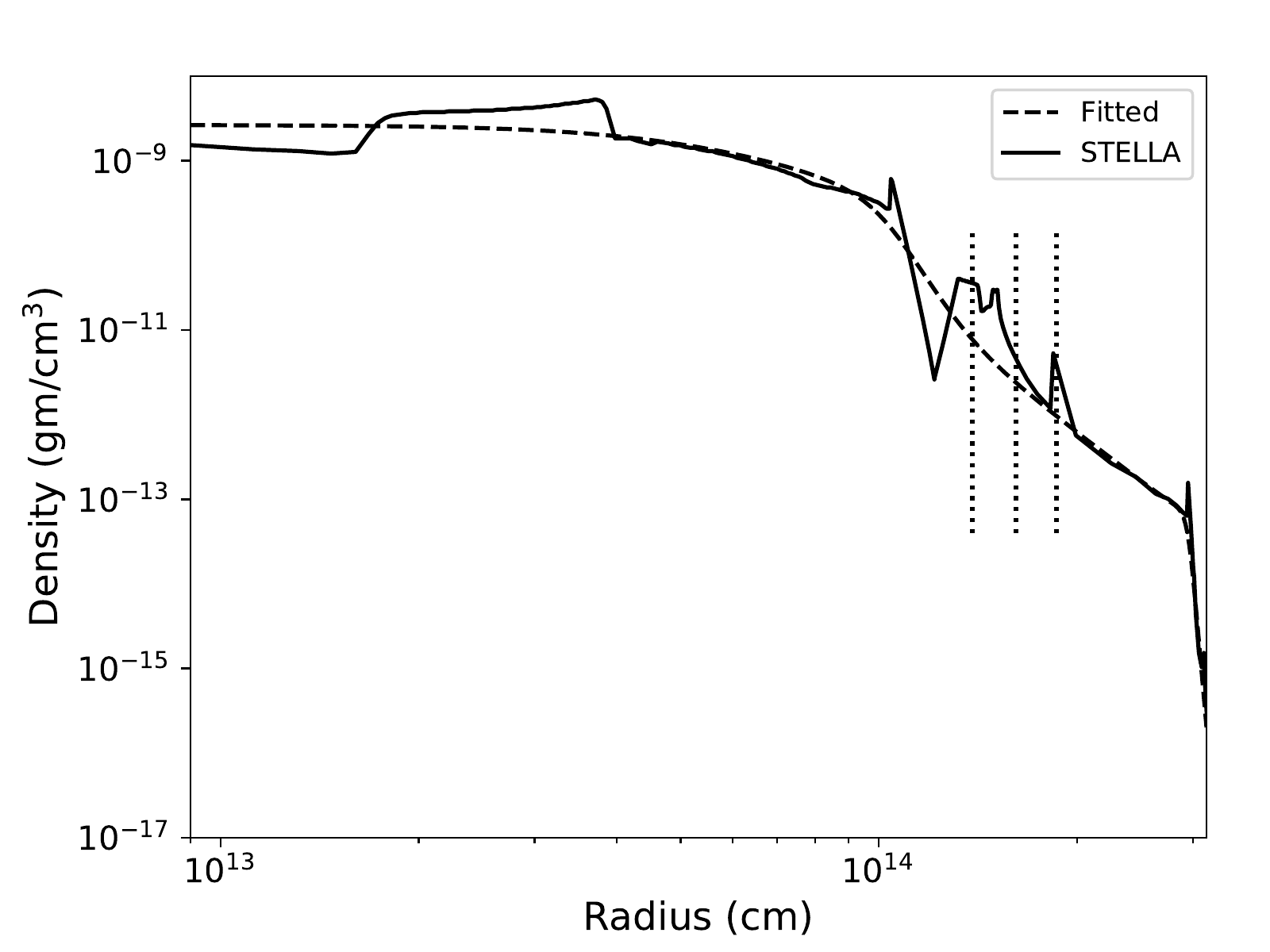}
\caption{Ejecta structure of SN 1993J (left panel) and SN 2011dh (right panel) at day 1 after the explosion of respective supernovae from STELLA. The fitted ejecta profiles are shown in dashed lines. The dotted line from right to left display the positions of the reverse shock in the ejecta at 1000, 2000, 3000, 5000, 7000 and 8000 days since the explosion of SN 1993J  (left panel) and at 200, 500 and 1200 days after the explosion of SN 2011dh (right panel).}
\label{fig:ejectastruc}
\end{figure*}

The ejecta profiles we use in our simulations are taken from STELLA, which is a multi-group radiation hydrodynamics code \citep{blinnikov98,blinnikov00,blinnikov06}. The ejecta structures at day 1 after the explosion are shown in solid lines in Figure \ref{fig:ejectastruc}, which are re-scaled, considering homologous expansion, from the ejecta profiles calculated at 490 and 462 days after the outbursts of SN~1993J and SN~2011dh, respectively. The stellar model considered in STELLA for SN 1993J and SN 2011dh are the re-mapped models 13C of \citet{woosley94} and 13Cdh, respectively. The remapping procedure for the STELLA grid is described in  \citet{blinnikov98}. The models were brought to the static equilibrium with preserving $P(\rho)$ structure identical to the original model 13C and keeping the total mass and the chemical composition for both models identical to 13C. Here $P$ and $\rho$ represent pressure and density, respectively.  13Cdh is derived from the 13C of \citet{woosley94}, where
         the radius of the pre-supernova star has been re-scaled to match the faster and weaker shock breakout maximum of SN 2011dh. The explosion energy of SN~2011dh that allow to fit the observations is $2\times 10^{51}$~erg, which is a little larger than that of SN~1993J. The detais of the model and the multicolor light curves are compared to the observations of SN~2011dh in \citet{tsvetkov12}.
%%The stellar model considered in STELLA for SN~1993J and SN~2011dh are the models 13C of \citet{woosley94} and 13Cdh, respectively. 13Cdh is derived from the 13C of \citet{woosley94}, where the radius of the pre-supernova star has been re-scaled to match the radius of the observed progenitor star of SN 2011dh.
%The stellar model considered in STELLA for SN 1993J is the model 13C of \citet{woosley94}, while for SN 2011dh the radius of the presupernova star in the 13C model has been re-scaled to match the radius of the observed progenitor star.   
The model 13C is a binary model in which the primary, with a main sequence mass of 13 $\msun$, has lost most of its hydrogen envelope to a nearby companion star of mass 9 $\msun$. The initial separation between the two stars was 4.5 AU. As the primary undergoes large mass loss because of strong interaction with the secondary, around 0.2 $\msun$ of the hydrogen envelope is left just before explosion. The pre-supernova star has a radius of $4.33 \times 10^{13}$ cm and the core, composed of helium and heavy elements, has a mass of 3.71 $\msun$. It's interesting to note that the density structure here is different from what is shown in Figs.3, 4 by \citet{blinnikov98} for the age 1 day after the explosion. This is a good illustration to the absence of the fully homologous expansion stage of the evolution for the SNe with a region of $^{56}$Ni inside. The thermalization of gamma-photons from $^{56}$Ni to $^{56}$Co to $^{56}$Fe decays leads to the formation of a hot bubble with the depressed density, which causes the formation of additional shocks and continues to change the gas distribution along the ejecta during several weeks after the explosion. The comparison between the hydrodynamical structure provided by radiation-hydrodynamical calculations and the Monte-Carlo calculations was discussed in \citet{woosley07}. For the present work, the details of $\rho(v)$ distribution also can be important, where $v$ represents velocity.

%The bumps visible in Figure~\ref{fig:ejectastruc} for both SNe are due to Ni bubble heating. More details about STELLA can be found in \citet{blinnikov98}.           

\section{Hydrodynamical simulation}
\label{sec:hydro_sim}
Often it is assumed that the interaction between SN ejecta and CSM can be described by a self-similar structure \citep{che82a}. This is a reasonable approximation if the outer part of the ejecta and CSM both have a power-law density profile, and the ejecta density profile is not too shallow. However, from multi-group radiation hydrodynamics simulation (e.g., STELLA) of an exploded star there is evidence that the density of the outer part of a SN can differ from a power-law structure. In this case the evolution of the SN can no longer be approximated by a self-similar solution, and one needs hydrodynamical simulations to estimate the structure. 

\par 
As displayed in Figure~\ref{fig:ejectastruc} (in solid lines), the ejecta profiles of both SNe~1993J and 2011dh, estimated from STELLA, are quite complicated. Therefore, we evaluate the shock structure by performing hydrodynamical simulation of SN and CSM interaction. To do the simulation we use the publicly available FLASH code, the Radiation Blast Wave problem, which has been modified accordingly to meet our requirements. FLASH \citep{fryxell00}{\footnote{ also see the FLASH manual at \\ http://flash.uchicago.edu/site/flashcode/user$\texttt\_$support/flash4$\texttt\_$ug$\texttt\_$4p4.pdf}} is a parallel, hydrodynamical code which is written in Fortran 90 and C. The hydrodynamical modules implemented here assume that the flow can be described by the Euler equations for compressible gas. The hydro unit is inherited from the PROMETHEUS code \citep{fryxell89} and uses the split Piecewise-Parabolic Method (PPM). PPM is described in detail in \citet{colella84}. To solve the Euler equations numerically a Riemann solver, known as the Harten, Lax and van Leer solver \citep{harten83}, is called in our simulations. We use a Courant-Friedrichs-Lewy (CFL) value of 0.6 for these runs.  
%The hydro unit we are using is inherited from the PROMETHEUS code \citep{fryxell89} and uses the split Piecewise-Parabolic Method (PPM)  

\par 
The original Radiation Blast Wave problem consists a gas, that is initially at rest and has a constant density. The temperature of the gas and radiation are not same inside a sphere of radius $R_{\rm int}$, centered at the origin. However, they are in thermal equilibrium beyond $R_{\rm int}$. In our case the gas is  made up of 63 \% hydrogen and 37\% helium (by mole fraction) and obey the ideal gas law, with an adiabatic index 5/3. Radiation is not included in our simulations. The temperature of the gas is 100 K and the density profile is a combination of the density structures of SN ejecta and CSM. In  Figure~\ref{fig:ejectastruc} the density of the ejecta, at day 1 after the explosion, are shown for both SN~1993J (left panel) and SN~2011dh (right panel). The radius of the ejecta at this epoch is $R_{\rm int} \equiv 2.24 \times 10^{14}$ cm ($3.15 \times 10^{14}$ cm) for SN~1993J (SN~2011dh). We assume that our SNe start to interact with the ambient media one day after the stars go off. Therefore, in our simulation the input density consists the density profile of ejecta ($\rho_{\rm ej}$) up to a radius $R_{\rm int}$, and beyond this the density profile is given by the CSM ($\rho_{\rm csm}$) such that at day 1, just prior to the interaction, $\rho_{\rm ej}(R_{\rm int}) = \rho_{\rm csm}(R_{\rm int})$.  For SN 1993J (SN 2011dh) the density structures of CSM is given by eq. \ref{eq:csm2} (winds, see \S \ref{subsubsec:sn1993j} for details). The STELLA outputs, i.e., the ejecta structures, are fitted with analytic functions to make the initial conditions smooth. These analytic fits to the ejecta structures are shown in dashed lines in Figure~\ref{fig:ejectastruc}.

\par
\begin{figure}
\centering
\includegraphics[width=8cm,angle=0]{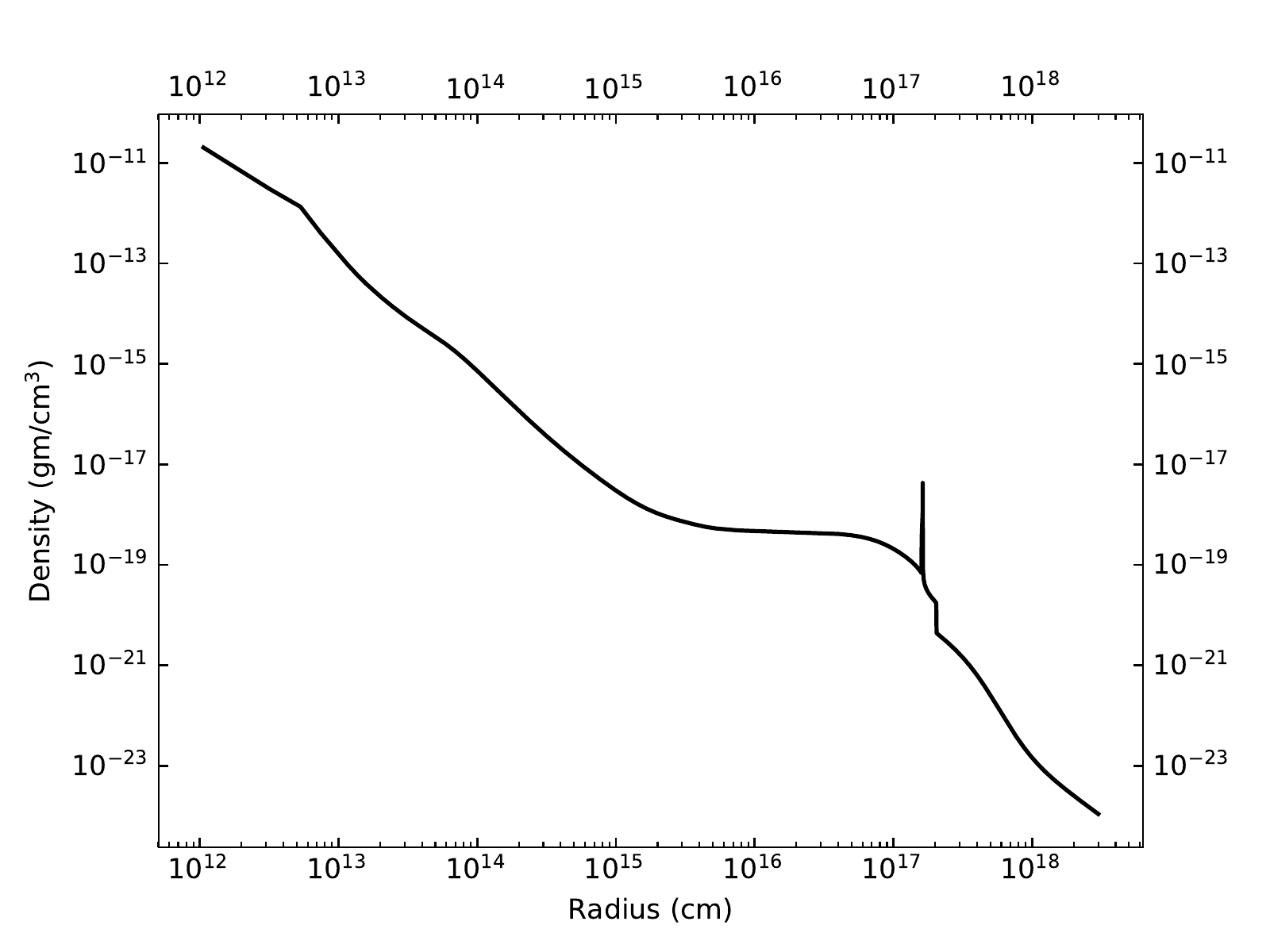}
\caption{Density profile of the expanding SN ejecta with the shocked gas and unshocked CSM post 2000 days since the explosion of SN~1993J. Note the steep slope in the unshocked CSM beyond $2\EE{17}$ cm, which then rolls off to an $r^{-2}$ wind at $r \geq 2\EE{18}$ cm.}
\label{fig:93JdenTotal}
\end{figure}

\begin{figure*}
\centering
\includegraphics[width=8cm,angle=0]{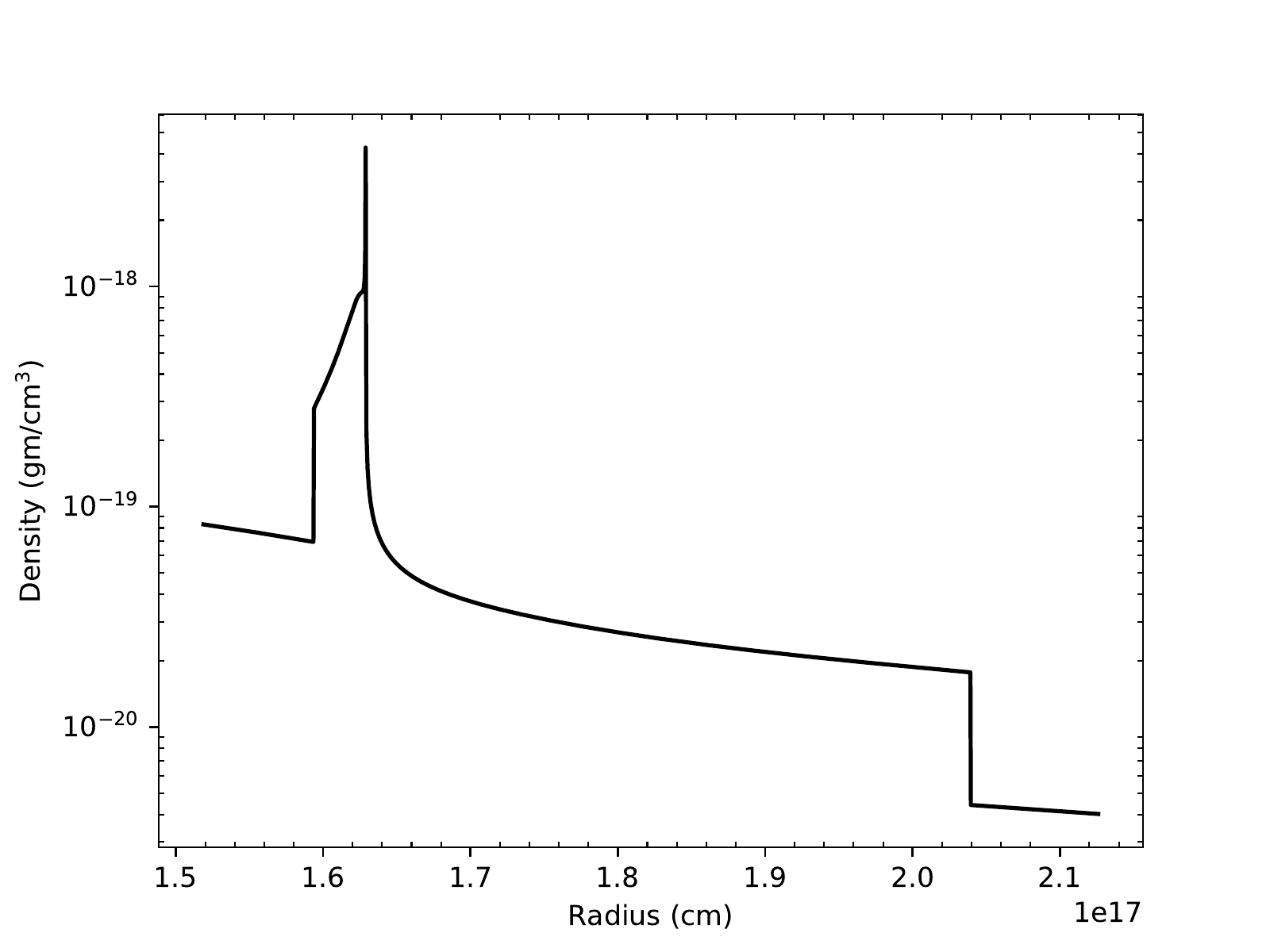}
\includegraphics[width=8cm,angle=0]{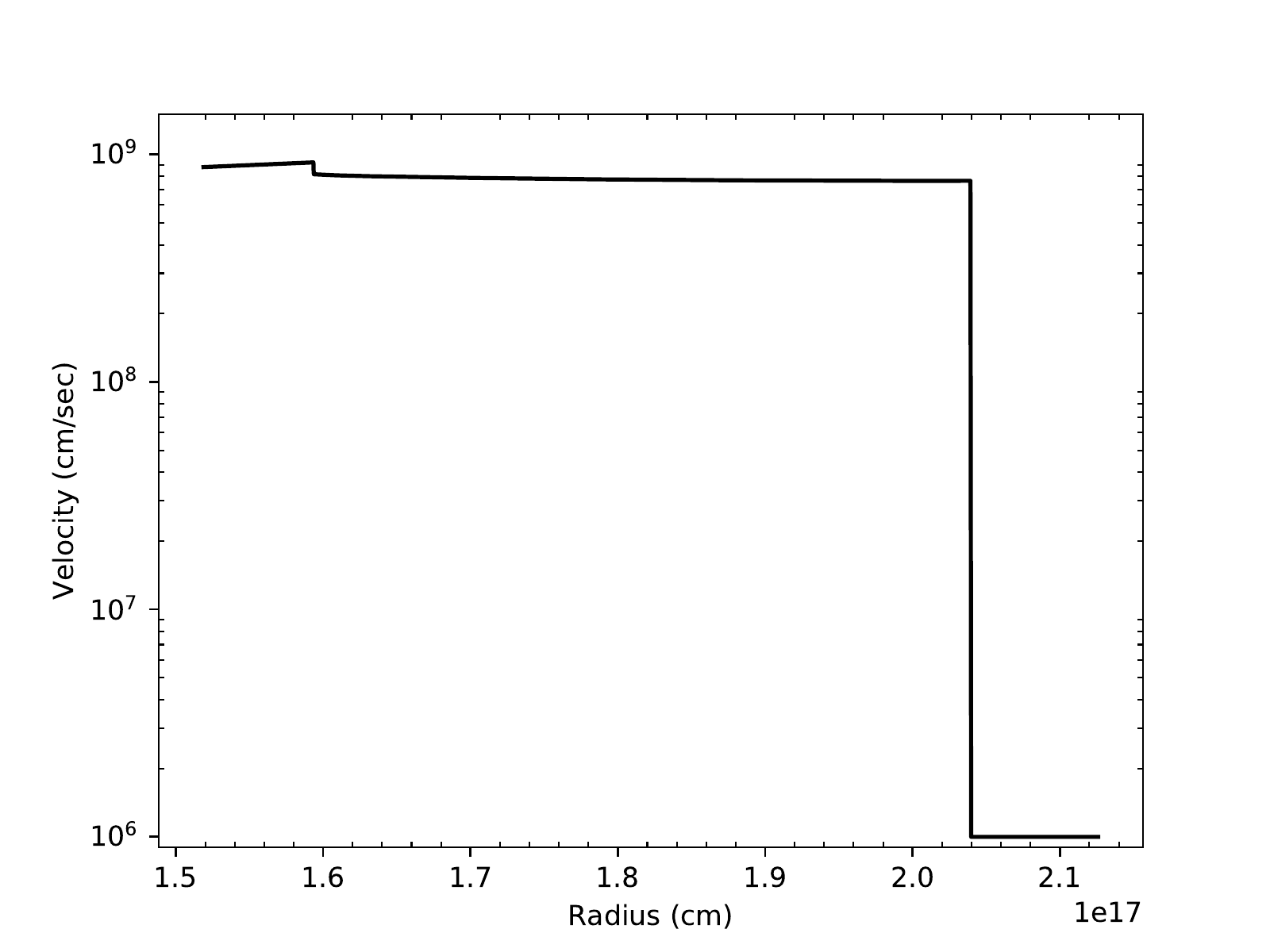}
\includegraphics[width=8cm,angle=0]{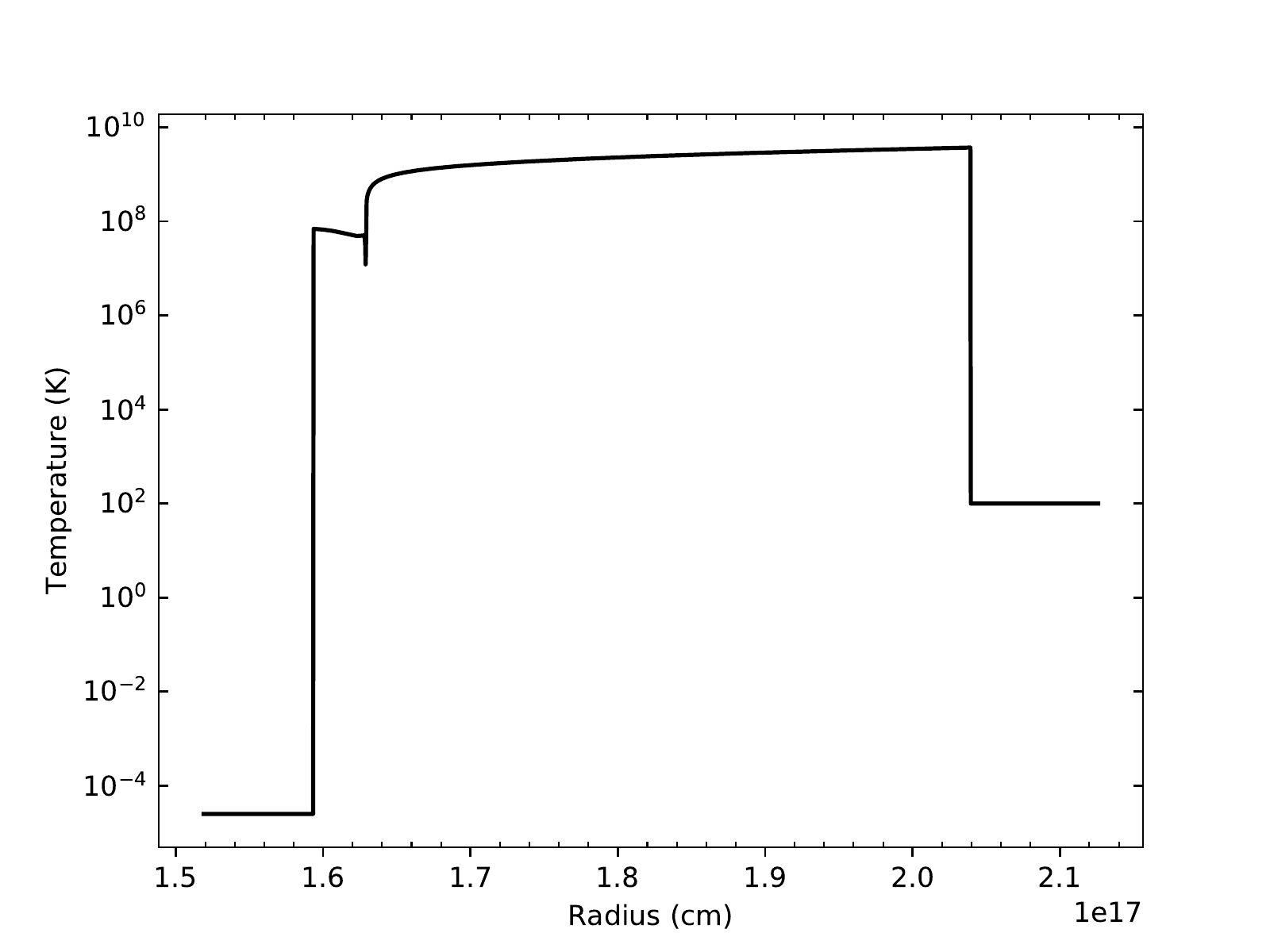}
\includegraphics[width=8cm,angle=0]{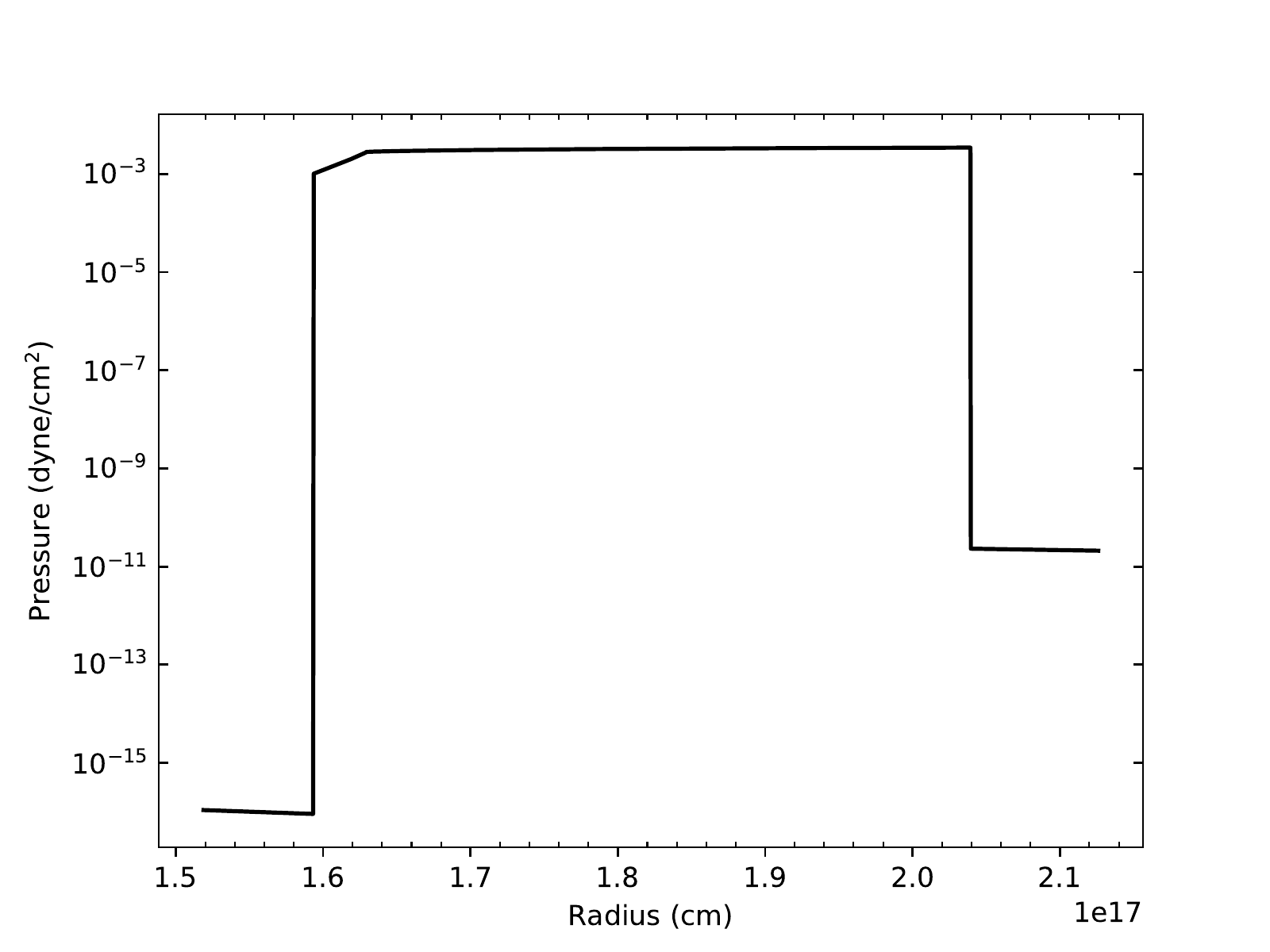}
\caption{Morphology of density (upper left panel), velocity (upper right panel), temperature (lower left panel) and pressure (lower right panel) across the shocked region at 2000 days post the explosion of SN 1993J.}
\label{fig:93Jshockprofile}
\end{figure*}

\begin{figure}
\centering
\includegraphics[width=8.5cm,angle=0]{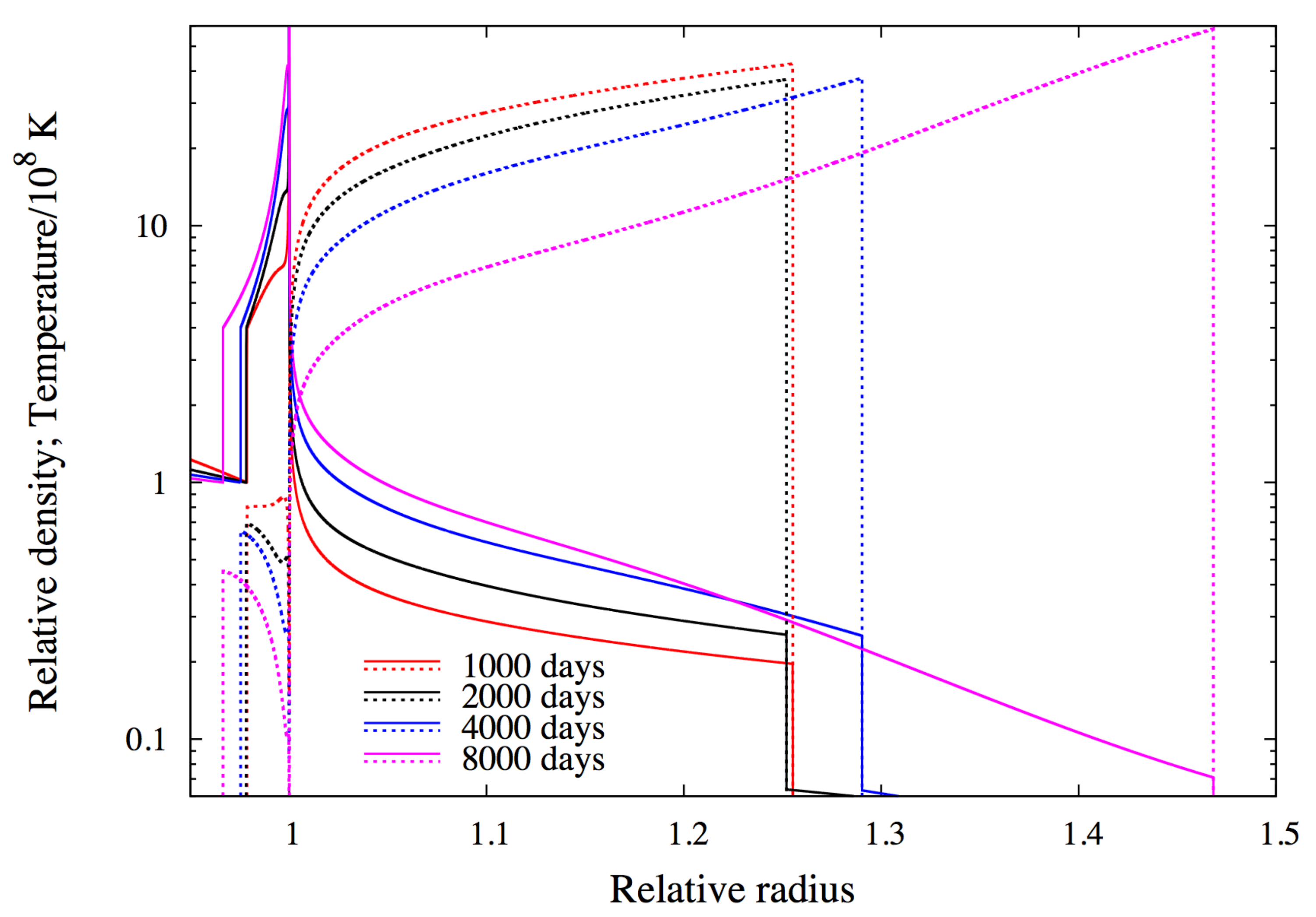}
\caption{Density (solid lines) and temperature (dotted lines) profiles across the shocked region at 1000, 2000, 4000 and 8000 days past the explosion of SN 1993J. Note the inflated region of shocked CSM at the latest epoch, and the increasing shock temperature with time of the forward shock.}
\label{fig:93JdenSh357}
\end{figure}

 
The simulations are performed in one dimension considering spherical geometry. The computational domain for SN~1993J and SN~2011dh are 0 cm $< r < 3\times 10^{18}$ cm and 0 cm $< r < 1 \times 10^{18}$ cm, respectively, where $r$ represents radius. We use a uniform grid with each cell has a width of $2.13 \times 10^{12}$ cm ($6.51 \times 10^{11}$ cm) for SN 1993J (SN 2011dh).  
%Therefore each block has same resolution.          
The interaction of the supersonic SN ejecta with the almost stationary CSM launches two shocks; one moves forward into the CSM, hence called the forward shock, and the other recedes into the ejecta in mass coordinate, and is known as the reverse shock. We run our FLASH simulations until 8000 days (3000 days) after the explosion of SN~1993J (SN~2011dh). The FLASH output of the density profile of the entire simulation including both of the expanding unshocked SN ejecta and CSM is shown in Figure~\ref{fig:93JdenTotal} at 2000 days past explosion of SN~1993J, and a zoom in on the morphology of the density, velocity, temperature and pressure across the shocked region is shown in Figure~\ref{fig:93Jshockprofile}. To highlight the evolution, the density and temperature profiles of the shocked gas at 1000, 2000, 4000 and 8000 days after explosion of SN~1993J are displayed in Figure~\ref{fig:93JdenSh357}. Solid lines are for the density. 
%The FLASH output of the interaction along with the density profile of the expanding unshocked SN ejecta and CSM are shown in Figure~\ref{fig:93JdenTotal} at 2000 days past explosion of SN 1993J. The morphology of the shock variables density, velocity, temperatures and pressure, across the shock are shown in Figure~\ref{fig:93Jshockprofile}. 
In computing the evolution of the shocks no radiative cooling was included in the hydro code. Cooling is chiefly important at early epochs when the density of the shocked gas is higher, and inverse Compton scattering in the region close to the forward shock plays a role. Therefore, these processes will mainly affect the evolution at early epochs \citep{fra96}. At late time, radiative cooling will only have an insignificant effect, as discussed in \S \ref{sec:Xray_result}. Of potentially greater importance is that we have assumed full equipartition between electrons and ions. In reality, the electron temperature may be lower than the ion temepratures of the shocked gas. We discuss this further in \S \ref{sec:Xray_result}.

%\begin{figure}
%\centering
%\includegraphics[width=8cm,angle=0]{SN93J_density_1d_zoomed2_015.pdf}
%\caption{Density profile across the shocked region at 3000 days past the explosion of SN 1993J.}
%\label{fig:93JdenSh3000}
%\end{figure}

%\begin{figure}
%\centering
%\includegraphics[width=8cm,angle=0]{SN93J_density_1d_zoomed2_025.pdf}
%\caption{Density profile across the shocked region at 5000 days past the explosion of SN 1993J.}
%\label{fig:93JdenSh5000}
%\end{figure}

%\begin{figure}
%\centering
%\includegraphics[width=8cm,angle=0]{SN93J_density_1d_zoomed2_035.pdf}
%\caption{Density profile across the shocked region at 7000 days past the explosion of SN 1993J.}
%\label{fig:93JdenSh7000}
%\end{figure}

\section{Radio and X-ray data}
\label{sec:data}
 In this paper, for SN~1993J, we have modeled the post 1000 days radio data published in \citet{weiler07} (for details see the references therein), \citet{dwa14}, as well as that presented in Das et al. (in preparation), while for the X-ray modeling we used data from \citet{chandra09}. Furthermore, for completeness we include the X-day data presented in \citet{dwa14}.
 %, however, these data are not considered for modeling ({\bf Peter: If you think we should mention something more here regarding Vikram's X-ray data please write.}). 
 As we are interested in modeling the late time emissions from both SNe, for SN~2011dh, we compare our model's prediction with the X-ray observation carried out $\sim$ 500 days after the explosion \citep{maeda14}. 
 %For radio modeling the data published in \citet{sod12} and \citet{krauss12} are considered ({\bf Peter : add the other reference(s) of publicly available radio data for SN 2011dh that were used to model the emission.}). 
 For radio we reanalyzed the publicly available data taken with Karl G. Jansky Very Large Array (VLA\footnote{The VLA is operated by the National Radio Astronomy Observatory, a facility of the National Science Foundation operated under cooperative agreement by Associated Universities, Inc.}), from the surroundings of SN 2011dh, taken on 1 Aug 2012, 31 Jan 2014 and 18 Oct 2014. Along with these data published in \citet{sod12}, \citet{krauss12} and \citet{deWitt16} are considered to model the radio emission. The distances to SN 1993J and SN 2011dh considered in our study are 3.63 Mpc \citep{freedman94} and 8.4 Mpc \citep{feldmeier97}, respectively.
 
 %We downloaded publicly available data taken with the Karl G. Jansky Very Large Array (VLA\footnote{The VLA is operated by the National Radio Astronomy Observatory, a facility of the National Science Foundation operated under cooperative agreement by Associated Universities, Inc.}), from the surroundings of SN 2011dh, taken on 1 Aug 2012, 31 Jan 2014 and 18 Oct 2014.  
\par
\begin{figure}
\centering
\includegraphics[width=8.5cm,angle=0]{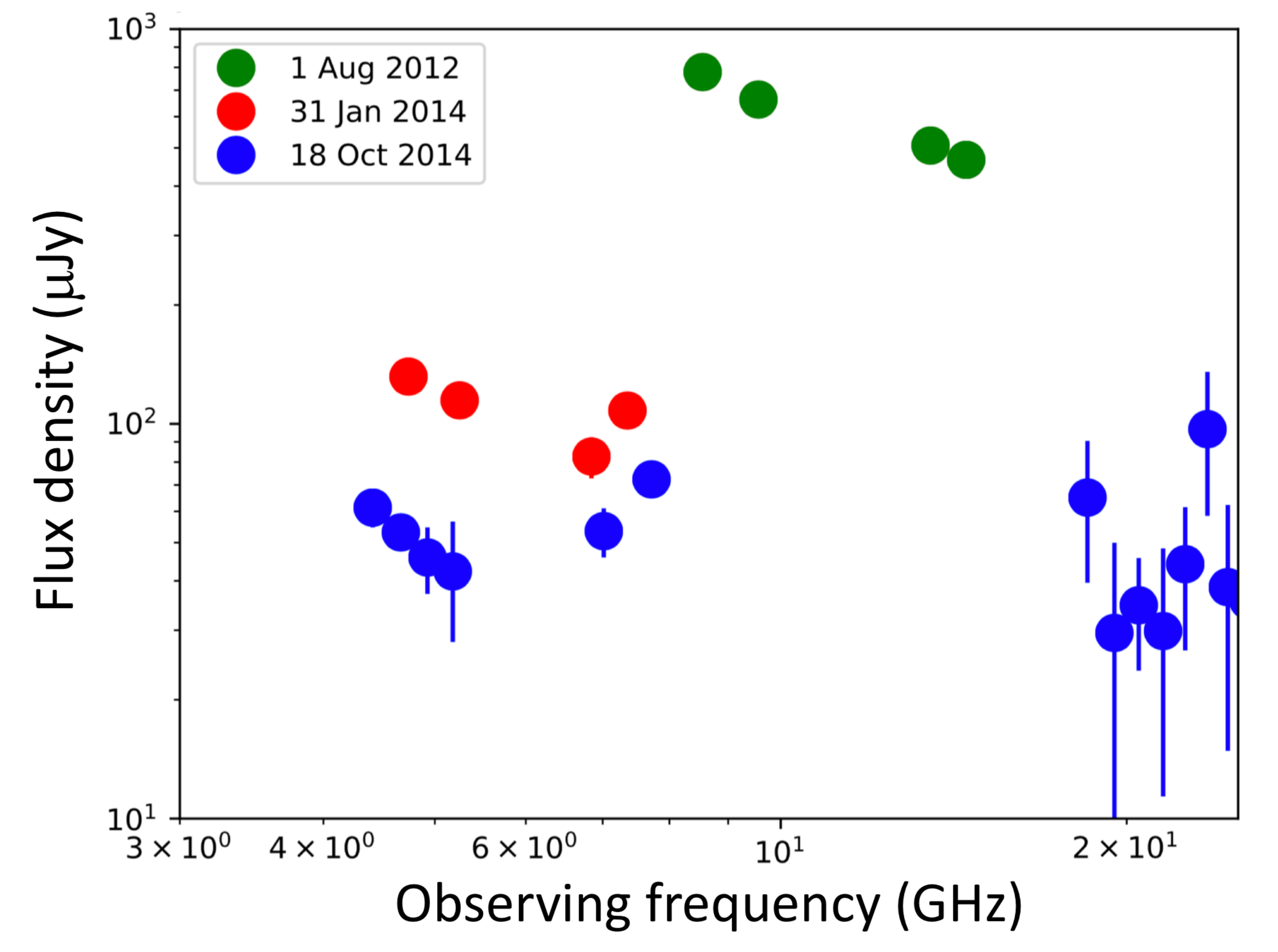}
\caption{Broad-band VLA radio spectra of SN 2011dh between 2012 and 2014. 
%Note a similar increase in flux density at around 8 GHz in the data acquired on 31st January and 18th October 2014.
} 
\label{SN2011dh-VLA}
\end{figure}

 
The observations presented here were obtained as part of the VLA programs 12A-286, 13A-370, and 14A-479.  We processed and imaged the data using the Common Astronomy Software Application (CASA) software package \citep{mcmullin07}, The source 3C286 was used in all three observing runs for bandpass and absolute flux density calibration purposes, for all frequencies and epochs. For phase calibration purposes, the sources J1335+4542, J1327+4326, and J1349+5341 were used for the 1 Aug 2002, 31 Jan 2014, and 18 Oct 2014 observations. Our results are summarized in Table \ref{VLA-log}, and in Figure~\ref{SN2011dh-VLA}. We note that our 8.4 GHz data point for our 31 Jan 2014 (PI: Arcavi) epoch indicates a flux density about three times smaller than the flux density published
by \citet{deWitt16}. Since this large difference in flux density has a correspondingly large impact in the radio light curve and its physical interpretation, we carefully re-analyzed all three data sets, in particular the
data for 31 Jan 2014, as there were many sources in the field surrounding the SN. After this procedure, we confirmed that the source position for the VLA source that we identify with SN~2011dh in both 31 January and 18 October 2014,
agrees within about 1 milliarcsecond with the VLBI position of SN~2011dh published by \citet{Mart11b} and with the position published by \citet{deWitt16}. We note that there is a source northwards of the actual SN~2011dh, which has a flux density of about 600 $\mu$Jy/beam, i.e., about ten times larger than SN~2011dh.  We also notice that the observations of 18 October 2014 (PI: Arcavi) were phase-centered to a position significantly off the SN discovery position.  Therefore, we suspect that \citet{deWitt16} might have been confused by that source northern of the SN, which is about ten times brighter than the actual radio luminosity of SN~2011dh for this epoch. 

\par 
The broad-band VLA data displayed in Figure ~\ref{SN2011dh-VLA} shows three
different epochs, and an inverted spectral behavior at frequencies $\geq$5.0 GHz
both in January and October 2014. 

\begin{deluxetable*}{crrr}
\tablecolumns{4}
\tablewidth{0pt} 
\tablecaption{Karl G.~Jansky Very Large Array Radio Observations of SN 2011dh }
\tablehead{
	\colhead{Epoch} & 
	\colhead{Day since explosion} &
	\colhead{Frequency} & 
	\colhead{Flux Density} \\ 
	\colhead{yyyy-mmm-dd} &
	\colhead{days} &
	\colhead{[GHz]} &
	\colhead{[$\mu$Jy]}
}
\startdata
2012-Aug-01 & 428 &  8.4  & 880$\pm$60\\
            &     &  8.55  & 778$\pm$24\\
            &     &  9.56  & 663$\pm$23\\
            &     &  13.50  & 507$\pm$17\\
            &     &  14.50  & 466$\pm$17\\
2014-Jan-31 & 976 &  4.74  & 132$\pm$6\\
            &     &  5.26  & 115$\pm$8\\
            &     &  6.84  & 83$\pm$9\\
            &     &  7.36  & 108$\pm$3\\
            &     &  8.4   & 180$\pm$20\\
2014-Oct-18 & 1236  &  4.42  & 61$\pm$6\\
            & &  4.67  & 53$\pm$5\\
            & &  4.93  & 46$\pm$9\\
            & &  5.18  & 42$\pm$14\\
            & &  7.02  & 53$\pm$8\\
            & &  7.72  & 72$\pm$5\\
            & &  8.4   & 95$\pm$20\\
            & &  18.49 & 65$\pm$25\\
            & &  19.51 & 30$\pm$21\\
            & &  20.49 & 35$\pm$11\\
            & &  21.51 & 30$\pm$18\\
            & &  22.49 & 44$\pm$17\\
            & &  23.51 & 97$\pm$38\\
            & &  24.49 & 39$\pm$24\\
            & &  25.51 & 36$\pm$26
\enddata
%\tablecomments{}
\label{VLA-log}
\end{deluxetable*}

\section{Radio emission}
\label{sec:radio_emission}
Radio emission originates from the shocked gas enclosed between the reverse and the forward shocks. The forward shock interacts with the immediate surrounding medium and accelerates the charged particles in the ambient medium  in relativistic energies.  These shocks are also ideal places where effective magnetic field amplification takes place \citep{bykov13, cap14a,cap14b,kundu17}. The shock accelerated charged particles in the presence of the magnetic field emit part of their energy as synchrotron radiation in the radio wavelengths. According to diffusive acceleration theory the energy distribution of the accelerated particles follow a power-law profile, i.e, $dN/dE=N_0 E^{-p}$, where $N_0$ and $E$ are the normalization constant and energy of particles, respectively. $p$ represents the power law index. It is very likely that in the acceleration process the seed particles acquire a certain percentage of the post shock energy.
As electrons over ions are expected to be the main source of radio emission in SNe, we assume that in the post-shock region all electrons are accelerated, and that a constant fraction of post shock energy, $\epsilon_{\rm e}$, gets channeled into them. Here  $\epsilon_{\rm e} = u_e/u_{th}$, where $u_e$ represents the energy density of electrons. $u_{th} = 9/8 \rho_{\rm csm}(r) v_s^2$ is the post shock thermal energy density with $v_s$ being the shock speed.
%Here  $\epsilon_{\rm e} = u_e/u_{th}$, where $u_e$ and $u_{th}$ represent the energy density of electric field and post shock thermal energy density, respectively. 
%In a similar fashion 
As for electrons, we presume that the fraction of post-shock kinetic energy that is converted into magnetic field is $\epsilon_{\rm B} = u_B/u_{th}$, where $u_B$ is the energy density of the magnetic field. In modeling the radio emission it is surmised that the emission is coming from a spherical homogeneous shell. Therefore, considering synchrotron self-absorption (SSA) and external free-free absorption as the dominating absorption mechanisms, the optically thin luminosity of this radiation, due to relativistic shock accelerated electrons with a power-law index of $p$, from a shell having an outer radius and thickness of $r_s$ and $\Delta r$, respectively, can be written as 

\begin{equation}
	L_{\nu, {\rm thin}}=\frac{8 \pi^{2} k_B T_{\rm bright} \vartheta_{\nu} r_s^2} {c^2 f\left(\frac{{\nu}_{\rm peak}}{{\nu}_{\rm abs}}\right)}
	                 \nu_{{\rm abs},0}^{(p+3)/2} \nu^{-(p-1)/2} {\rm exp}[{-\tau_{\rm ff}(r_s)}],
\label{eq:Luminosity}
\end{equation}
with

\begin{equation}
	\nu_{{\rm abs},0}= \left(2 \Delta r ~ \varkappa(p) ~N_0 ~ B^{(p+2)/2}\right)^{2/(p+4)},
\label{eq:nuabs}
\end{equation}
and
\begin{equation}
	f (x) = x^{1/2} \left[1 - {\rm exp} \left( - x^{-\left(p+4\right)/2} \right) \right],
\label{eq:fx}
\end{equation}
where $T_{\rm bright}$ is the brightness temperature. $\vartheta_{\nu} = \frac{L_{\nu}}{L_{\nu,0}}$ with $L_{\nu,0}  =  4 \pi^2 r_{s}^{2} I_{\nu}(0)$. $I_{\nu}(0)$ is the intensity of radiation along the line of sight for which the path length of the radiation is equal to the thickness of the shock. ${\nu}_{\rm peak}$ is the peak frequency of radiation. $\tau_{\rm ff}$ is the free-free optical depth. $B$ represents magnetic field strength and $\varkappa(p)$ is the SSA coefficient. $k_B$ and $c$ represent the Boltzmann constant and speed of light in vacuum, respectively. 
%As the ambient media for SNe 1993J is  it is expected that the radio emission will suffer external free free absorption. 

For SN~1993J, we have assumed a CSM out to $2\EE{17}$~cm which is characterized by a mass loss rate of $\mdot$, of $4 \times 10^{-5}$ $\msunyr$ and a wind velocity $v_w = 10$ $\kms$. For such a high mass-loss rate, radio emission from this SN could suffer from external free-free absorption. 
We assume that the ambient media of both SNe, SN 1993J and SN 2011dh, are made up of hydrogen and helium, with solar abundances, and that it is completely ionized due to strong UV radiation from shock break out and X-ray emission from forward and reverse shocks \citep{fra96}. If the number density of hydrogen and helium in the surrounding medium are $n_{\rm H}(r)$  and  $n_{\rm He}(r)$, respectively, the absorption coefficient due to free-free is given by 
\begin{equation}
\begin{split}
\alpha_{\rm ff}(r) = 0.018 ~ T_{\rm csm}^{-3/2} ~ \frac{n_{\rm e}^2(r)}{1+2y} ~\nu^{-2} ~ \big[{\bar g}_{\rm ff}^{\rm H}(T_{\rm csm},\nu) ~ \\+ ~ 4y ~ {\bar g}_{\rm ff}^{\rm He} (T_{\rm csm},\nu)\big]~ ~{\rm cm}^{-1},
\end{split}
\label{eq:alphaCoefff} 
\end{equation}
where $T_{\rm csm}$ is the temperature of the CSM. Here $n_{\rm e}(r)$ represents the number density of the electron in the ambient medium and $y = n_{\rm He}(r)/n_{\rm H}(r)$. $ {\bar g}_{\rm ff}^{\rm H} (T_{\rm csm},\nu)$ and ${\bar g}_{\rm ff}^{\rm He} (T_{\rm csm},\nu)$ are the velocity average Gaunt factors for hydrogen and helium, respectively. In case of SN 1993J the temperature of the CSM is $\sim 10^5$ K \citep{bjornsson14}. For radiation in GHz frequency and $T_{\rm csm} < 3 \times 10^{5} z^2$, ${\bar g}_{\rm ff} (T_{\rm csm}, \nu) = \frac{\sqrt{3}}{\pi} \bigg( 17.7 + {\rm log} \big(\frac{T_{\rm csm}^{1.5}}{z \nu}\big) \bigg)$, where $z$ is the atomic number of a given element, while for $T_{\rm csm} > 3 \times 10^{5} z^2$ the Gaunt factor is given as ${\bar g}_{\rm ff} (T_{\rm csm}, \nu) = \frac{\sqrt{3}}{\pi}  {\rm log} \bigg(\frac{2.2 k T_{\rm csm}}{h \nu}\bigg)$ \citep{tucker75}. In case of a a wind medium, which extends up to a radius $r \gg r_s$,  $\tau_{\rm ff}$ is written as
\begin{equation}
%\tau_{\rm ff} = \frac{1}{3} n_e^2(r_s) r_s \alpha_{\rm ff}.
\tau_{\rm ff}(r_s) = \frac{1}{3} r_s \alpha_{\rm ff}(r_s).
\label{eq:tauff1}
\end{equation}
However, if the density of ambient medium starts to decrease, or increase, beyond a given radius $R_{\rm chng}$, the free-free optical depth becomes
\begin{equation}
%\tau_{\rm ff} = \frac{1}{3} n_e^2(r_s) r_s \alpha_{\rm ff} \bigg(1 - f^3_{{\rm R}_{\rm chng}} (1 -f^2_{\rm rate} )\bigg),
\tau_{\rm ff}(r_s) = \frac{1}{3} r_s \alpha_{\rm ff}(r_s) \bigg(1 - f^3_{{\rm R}_{\rm chng}} (1 -f^2_{\rm rate} )\bigg),
\label{eq:tauff2}
\end{equation}
where $f_{{\rm R}_{\rm chng}} = r_s/R_{\rm chng}$ and $f_{\rm rate} = \frac{\mdot^1/v^1_w}{\mdot/v_w}$. $\mdot^1/v^1_w$ is the ratio between the mass-loss rate and wind velocity which characterise the medium at $r \geq R_{\rm chng}$.

%\begin{equation}
%	I_{\nu}\left(h\right)=\frac{2kT_{\rm bright}} {c^2 f\left(\frac{{\nu}_{\rm peak}}{{\nu}_{\rm abs}}\right)}
%	                 \frac{\nu^{5/2}} {\nu_{{\rm abs},0}^{1/2}}
%	                 \left[
%	                 1-{\rm exp}\left(-\xi_h \tau_{\nu_{0}}\right)
%	                 \right],
%\label{eq:Intensity}
%\end{equation}
%with 
%\begin{equation}
%	f (x) = x^{1/2} \left[1 - {\rm exp} \left( - x^{-\left(p+4\right)/2} \right) \right]
%\label{eq:fx}
%\end{equation}
%\citep{bjo14}. Here $\xi_h  = \Delta s (h)/ (2 \Delta r)$ gives the normalised path length traversed by the radiation along the line of sight and $0 \leq h \leq 1$ takes into account emission from different parts of the shell. ${\nu}_{\rm abs}$ is the absorption frequency for which optical depth $\tau_{\nu_{\rm abs}} = 1$ and for $h = 0$ ${\nu}_{\rm abs} = {\nu}_{\rm abs,0}$ and $\tau_{\nu} = \tau_{\nu_{0}}$, respectively. $T_{\rm bright}$ and ${\nu}_{\rm peak}$ represent the brightness temperature and peak frequency of the radiation. $k$ is the Boltzmann constant. 

%According to the models, SNe~2011fe and 2014J are in the optically thin regime when the radio observations are done (see $\S$ \ref{sec:data} for radio observations). Therefore, the luminosity, for $\tau_{\nu} < 1$, from a shell with outer radius $r_s$ can be written as following :      
%
\section{X-ray emission}
\label{sec:Xray_emission}
We modeled the X-ray emission at late epochs, when the shocked ejecta behind the reverse shock mainly contributes to this emission.  The emission processes which we considered for our study are free-free, free-bound, two-photon and line emissions. In a medium where the external radiation is negligible the ionization structure of the shocked ejecta can be calculated by balancing collisional ionization with direct and dielectronic recombination. For an element $z$, if $X_{m}$ represents the ionization fraction of ionization stage $m$, then the rate of change of $X_{m}$ can be written as
\begin{equation}
\frac{dX_m}{dt} = N_e ~ [-(\beta_m + \mathbb{C}_m) X_m ~ + ~ \beta_{m+1} X_{m+1} ~ + ~ \mathbb{C}_{m-1}X_{m-1}], 
\label{eq:iostn_eq1}    
\end{equation}
\citep{nymark06} where $N_e$ is the number density of free electrons. $\beta_m$ represents the recombination coefficient from the ionization state $m$ to $m-1$ and $\mathbb{C}_m$ is the ionization rate from ionization state $m$ to $m+1$. We assumed steady state condition, i.e., $\frac{dX_m}{dt} = 0$, and solved the equation for each ionization stage of a given $z$, also including
an additional process, namely charge transfer. The elements 
included in our computer code to calculate the emission are H, He,
C, N, O, Ne, Na, Mg, Al, Si, S, Ar, Ca, Fe and Ni. 

\par 
In hot ejecta free electrons can excite atoms/ions to higher energy states, which can subsequently decay by the emission of photons. 
Collisional de-excitation is also important when the density of free electron is high enough. In this kind of medium, for an ionization state $X_m$ of a given element, which is in an excitation level $i$, under equilibrium we write
\begin{equation}
\begin{split}
\sum_{j\neq i} N(z_m)_j N_e q_{ji} + \sum_{j>i}N(z_m)_j A_{ji} = \sum_{j\neq i} N(z_m)_i N_e q_{ij} \\ + \sum_{j<i} N(z_m)_iA_{ij},
\end{split}
\label{eq:lineemission1}    
\end{equation}
with 
\begin{equation}
\sum_{j} N(z_m)_j = N(z_m),
\label{eq:lineemission2}    
\end{equation}
and 
\begin{equation}
q_{ij} = \int_{0}^{\infty} \mathbb{V} ~ \sigma_{ij} ~ f(\mathbb{V}) ~ d\mathbb{V}.
\label{eq:lineemission3}
\end{equation}
Here $N(z_m)$ represents the density of ionization stage $m$ of an element $z$. $A_{ij}$ and $\sigma_{ij}$ are the transition probability and collisional cross section for the $i$ to $j$ transition, respectively. $f(\mathbb{V})$ represents the thermal distribution of speed of free electrons with $\mathbb{V}$ being the speed of electrons. 

Recombination results in continuum free-bound emission, followed 
by a cascade of lines \citep{osterbrock89}. For thermal electrons,
the free-bound emission coefficient can be written as \citep{osterbrock89}
\begin{equation}
j_{\nu}^{fb} = \frac{1}{4\pi} ~ N(z_m) ~ N_{\rm e} ~\sum_{n=n_1}^{\infty} ~ \sum_{L=0}^{n-1} ~ \mathbb{V} ~ \sigma_{nL} ~ f(\mathbb{V}) ~ h ~ \nu ~ \frac{d\mathbb{V}}{d \nu}.
\label{eq:freebound2}    
\end{equation}
Here $\sigma_{nL}$ and $L$ represent the recombination cross section and orbital angular momentum, respectively. $h$ is the Planck constant.

Another continuum emission process is free-free emission, or bremsstrahlung. For  isotropic thermal electrons, that can scatter on an element $z$ in an ionization state $X_m$, the emission coefficient for free-free radiation is given as \citep{ryb79,osterbrock89}
\begin{equation}
\begin{split}
j_{\nu}^{ff} = \frac{1}{4\pi} ~ \frac{2^5 \pi e^6}{3 m_e c^3} ~ \bigg(\frac{2 \pi}{3 k_B m_e} \bigg)^{1/2}  ~ T^{-1/2} ~ z^2 ~ N(z_m) ~ N_{\rm e} \\ \times ~ exp(-h\nu/k_B T) ~ {\bar g}_{\rm ff} (T, \nu), 
\end{split}
\label{eq:freefree}    
\end{equation}
where $T$ and $m_e$ represent the temperature and rest mass of the electrons, respectively. $e$ is the electric charge.  

Continuum emission is also produced by two-photon decay, 
since electrons populated in the $2^2S$ level can decay to $1^2S$ by the emission of two photons. The emission coefficient corresponding to this radiation is
\begin{equation}
j_{\nu}^{2\gamma} = \frac{1}{4\pi} ~ N_{2^2S} ~ A_{2^2S,1^2S} 2 ~ h ~ y P(y), 
\label{eq:twophoton1}    
\end{equation}
\citep{osterbrock89} where in each decay, $P(y)$ is the normalized probability for one photon to emit in the energy range $y \nu_{12}$ and $(y+dy)\nu_{12}$. $h\nu_{12}$ is the energy difference between the two levels $1^2S$ and $2^2S$. $A_{2^2S,1^2S}$ and $N_{2^2S}$ represent the transition probability and density of electron in the $2^2S$ state, respectively.

\par
To calculate the X-ray emission, we have post-processed the output data from the adiabatic hydrodynamic models. The total emissivity due the above-mentioned processes are calculated using a plasma code which is described in some detail in \citet{soro04}, and was also used in \citet{matt08}. It is assumed in our model that the plasma is in collisional equilibrium for a given electron temperature. 

%To calculate the X-ray emission we have post-processed the output data from the adiabatic hydrodynamic models. This is done using a plasma code which calculates the ionization and emissivity as a function of density, temperature and composition. All elements included in the explosion models (up to Ni) are considered for the X-ray emission, and all ionization stages of the elements are included, as well as all important types of emission (i.e., free-free emission, recombination emission, two-photon emission and line emission). The plasma code is described in some detail in \citet{soro04}, and was also used in \citet{matt08}. The plasma is assumed to be in collisional equilibrium, and electrons and ions are in equipartition.

\section{Results}
\label{sec:results}
Both SN 1993J and SN 2011dh have been followed extensively in radio and X-ray wavelengths \citep{van93,pooley93,marca95a,marca95b,marca97,bartel00,PTorres01,bartel02,PTorres02,chandra04,weiler07,marca09,Mart11a,Mart11b,kohmura94,immler01,uno02,swartz03,zimm03,chandra05,chandra09,sod12,krauss12,deWitt16,bietenholz12,campana12,maeda14}
%citep{perezTorres01}
. Here we have modelled the radio and X-ray emission at late stages, post 1000 days for the former, and beyond $\sim$ 100 days for the latter, to gain information about the evolution of the progenitor before explosion, as well as the structure of the SN ejecta. 

\subsection{Radio}
\label{subsec:radio}
\subsubsection{SN 1993J}
\label{subsubsec:sn1993j}
\begin{figure}
\centering
\includegraphics[width=8cm,angle=0]{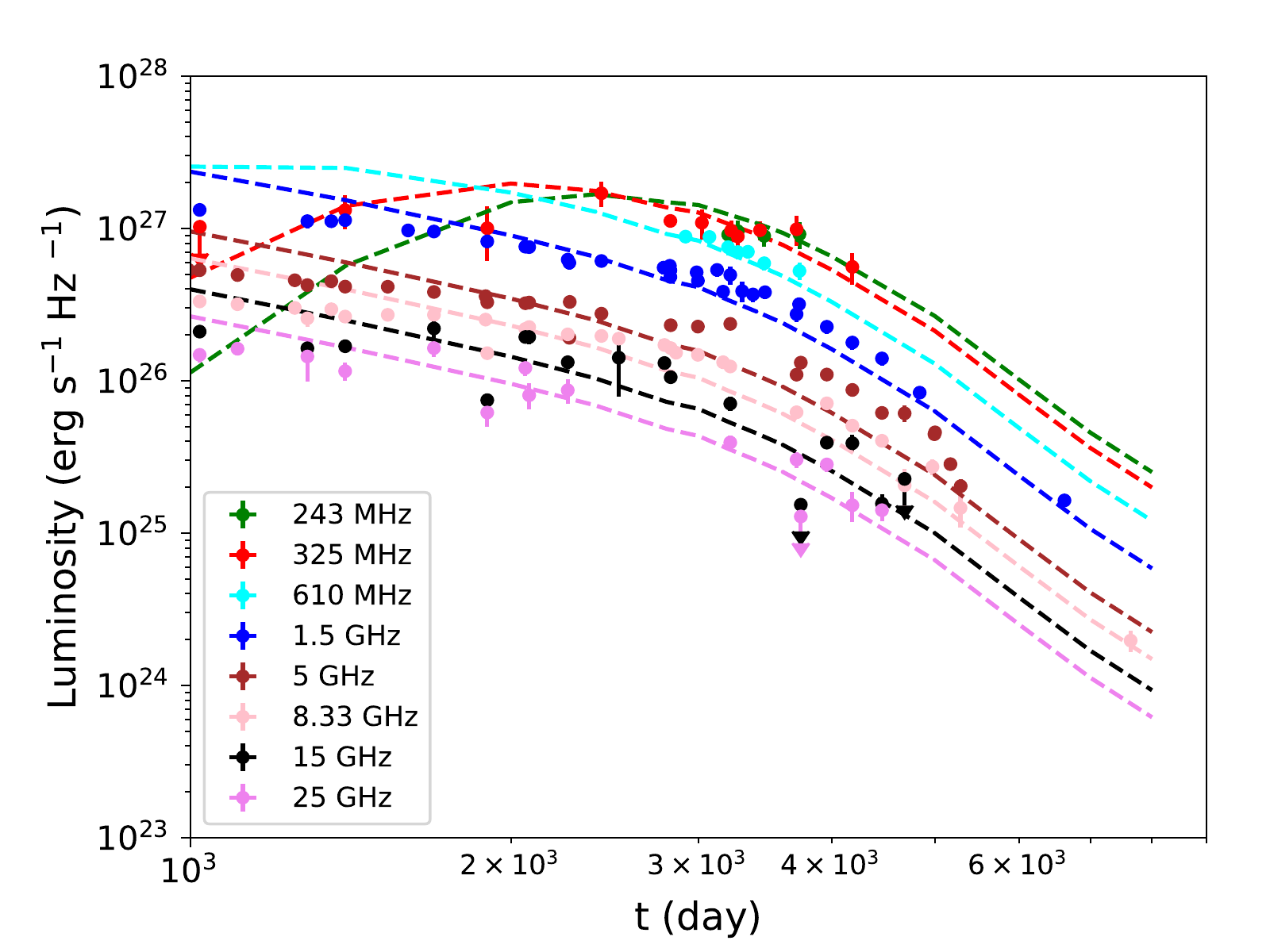}
\caption{Modeled radio light curves of SN 1993J post 1000 days. The data are from \citet{weiler07}, \citet{dwa14} and Das et al. (in preparation). The observed light curves show a sudden downturn in luminosity beyond $\sim$ 3000 days. From our analysis it is found that this drop is related to a rapid decrease in CSM density beyond a radius $2\EE{17}$ cm. The density of the ambient medium in our model is given by eqn.\ref{eq:csm2}. It is found that up to $\sim$ 3000 days the radio fluxes are consistent with a wind medium characterized by a $\mdot$ of $4 \times 10^{-5}$ $\msunyr$ and wind velocity of $v_w = 10$ $\kms$.}
%The ambient medium up to this radius is due to winds and can be characterized by a $\mdot$ of $4 \times 10^{-5}$ $\msunyr$ and a wind velocity $v_w = 10$ $\kms$. }
\label{fig:lightcurve93J}
\end{figure}

The radio light curves for SN~1993J at various frequencies are shown in Figure~\ref{fig:lightcurve93J}. The observed light curves show a sudden drop in flux beyond $\sim$ 3000 days. 
%From our radio modeling we found that before 3000 days the
%emission is consistent with a wind-like CSM, which is
%characterized by a mass loss rate $\mdot = 3.9 \times 10^{-5}$ $\msunyr$ 
%with a wind velocity of 10 $\kms$, while 
To replicate the sudden downturn in flux, it has been argued that this indicates a rapid decrease in the density of the surrounding medium at the position of the forward shock at $\sim 3000$ days. It is also possible that this downturn is related to fact that the reverse shock $\sim$ 3000 days post outburst starts to invade into the flat inner ejecta. This will affect the  evolution of the shocked gas encapsulated between the reverse and forward shock. 
%We did a careful examination of the the evolution of the shocks
%$\sim$ 3000 days and 
For the $\mdot / v_w$ we have assumed, and the explosion model used, we do not find that around 3000 days the reverse shock encounters any drastic change in the density profile of the unshocked ejecta gas. 
 
From our analysis it is found that the wind medium is extended up to a radius, $R_{\rm chng}$, of $2 \times 10^{17}$ cm. Beyond this the density starts to drop rapidly. To reproduce the observed radio light curves beyond 1000 days the density of the CSM in our model follows
\begin{equation}
%\rho_{\rm csm}(r) = \frac{\mdot}{4 \pi r^2 v_w} ~ \bigg(0.05 + \frac{0.95}{1 + \big(\frac{r}{4 \times 10^{17} {\rm cm}} \big)^4}  \bigg) ~~ \frac{\rm g}{\rm cm^3}
%\rho_{\rm csm} =   \rho^1_{\rm csm}/(1 + \big(\frac{r}{r_{\rm cut}}\big)^4 ),  
\rho_{\rm csm}(r) =   \rho^1_{\rm csm}(r) ~ ~ \bigg(0.05 + \frac{0.95}{1 + \big(\frac{r}{4 \times 10^{17} {\rm cm}} \big)^4}  \bigg) ~~\frac{\rm g}{\rm cm^{-3}}, 
\label{eq:csm2}
\end{equation}
where $\rho^1_{\rm csm}(r) = \mdot/4 \pi r^2 v_w$ with  $\mdot = 4\EE{-5}$ $\msun$ and $v_w =$ 10 $\kms$.
%and $r_{\rm cut} = 4 \times 10^{17}$ cm. 
The mass-loss rate and wind velocity considered in our model for $t < 3000$ days are similar to that required by \citet{fra96,fra98} to model the radio and X-ray emissions at early epochs. 

\par 
In Figure~\ref{fig:93JdenTotal} the rapid drop in CSM density is visible at $r >  2 \times 10^{17}$cm. Note that here we assume that the sudden drop in density continues up to a radius, $R_{\rm out}$, $2 \times 10^{18}$ cm, beyond which the CSM again attains a wind profile, characterized by a mass-loss rate of $\mdot/20$ and $v_w = 10$ $\kms$.  The radio 
%luminosity expected from 
 light curves computed from our model using 
an ambient medium given by eqn.\ref{eq:csm2} are shown in dashed lines in Figure~\ref{fig:lightcurve93J}. As shown in the figure, our calculated radio fluxes at different frequencies can reproduce the observed downturn with reasonable accuracy. It should be noted that in our model we did not include any local density fluctuations in the CSM. These variations could be the reason of the local fluctuations seen in the data.      

%\bf Very Long Baseline Interferometry (VLBI) images of SN1993J \st{taken at 8.4 GHz from day 50 until $\sim$ 2000 days} show\st{ed} that the radio emission \bf{\st{is coming} comes} from a nearly spherical shell  \citep{marca95a,marca95b,marca97,bartel00,marca09,Mart11b}, whose shell width is estimated to be $\sim$31\% of the external radius \citep{marca09,Mart11b}}

Very Long Baseline Interferometry (VLBI) images of SN1993J show that the radio emission comes from a nearly spherical shell \citep{marca95a,marca95b,marca97,bartel00,marca09,Mart11b}, whose shell width is estimated to be $\sim$31\% of the external radius \citep{marca09,Mart11b}. Therefore, in our calculation we assumed that for both SN~1993J and SN~2011dh the radiation comes from such a spherical shell. The evolution of this shell is calculated by performing hydrodynamical simulations, as described in $\S$ \ref{sec:hydro_sim}. At any given epoch the shell variables, e.g., the forward shock radius, velocity, density etc can be obtained from the simulations (see Figs.\ref{fig:93Jshockprofile} and \ref{fig:93JdenSh357}). The shock variables required for our calculations are radii of forward shock ($r_s$) and contact discontinuity ($r_c$), density of the shocked CSM ($\rho_{\rm sh,csm}$) and velocity of the forward shock ($v_s$). In upper left panel of Figure~\ref{fig:93Jshockprofile} the density spike on the left and the sharp discontinuity on the right side represent the position of $r_c$ and $r_s$, respectively. 
%Across the shock all variables suffer a discontinuity, that is why the cutoff is visible in other panel as well. 
We assume in our model (see \citet{kundu17} for details) that the radio emission is coming from the region between $r_c$ and $r_s$. Therefore, for us $\Delta r = r_s - r_c$. As displayed in Figs.\ref{fig:93Jshockprofile} and \ref{fig:93JdenSh357} there are variations in $\rho_{\rm sh,csm}$ and $v_s$ across $\Delta r$. Thus to get average values of these variables we determine them at $r = r_c + \Delta r/2 $. As can be seen in Figure 4, the density of the shocked CSM, except for close to $r_c$, is fairly constant with radius for the first few thousand days. Choosing parameters half-way through the shock is therefore not an important source of error. However, at 8000 days, there is a huge variation in density within the shocked CSM, so the error due to selecting representative parameters at this epoch is larger.

From spectral analysis, \citet{weiler07} found that both early and late epoch radio data are consistent with a spectral index of -0.8. Therefore, the power-law index, $p$, of the electrons responsible for this emission is 2.6. We did our modeling considering this value for $p$ and $T_{bright} = 3.6 \times 10^{10}$ K. As the radio source is expected to be homogeneous at late epochs the assumption of a constant value for $T_{bright}$ is reasonable. For other three parameters we assume $T_{\rm csm} = 6 \times 10^{5}$ K and $\epsilon_{\rm e} = \epsilon_{\rm B} = 0.03$.
% The other three parameters $T_{\rm csm}$, $\epsilon_{\rm e}$ have been determined from the fit. 
%It is found our study that to get a decent fit of the observed data we are required $T_{\rm csm} = 6 \times 10^{5}$ K, $\epsilon_{\rm e} = \epsilon_{\rm B} = 0.03$. 

\subsubsection{SN 2011dh}
\label{subsubsec:sn2011dh}
\begin{figure*}
\centering
\includegraphics[width=8cm,angle=0]{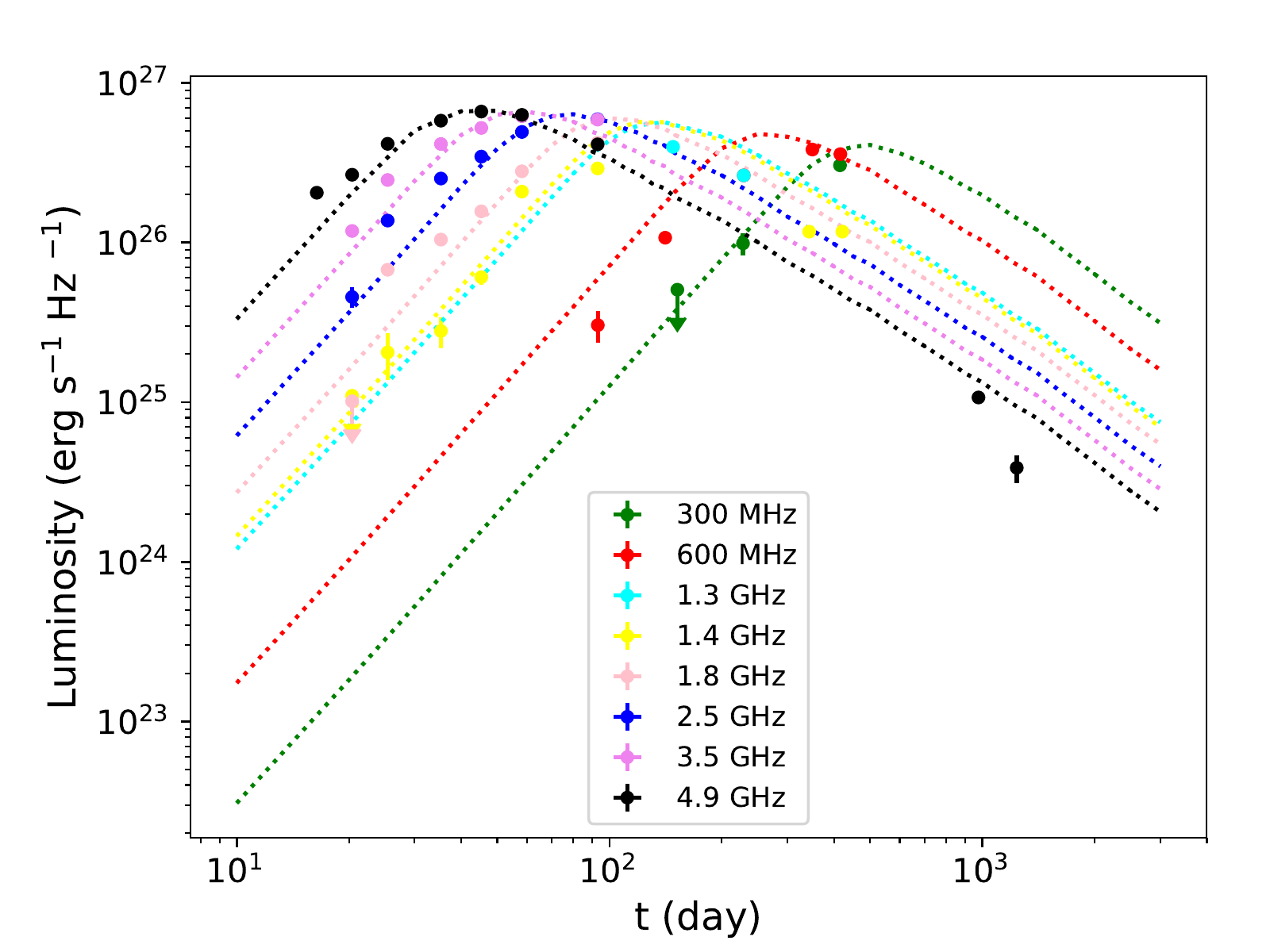}
\includegraphics[width=8cm,angle=0]{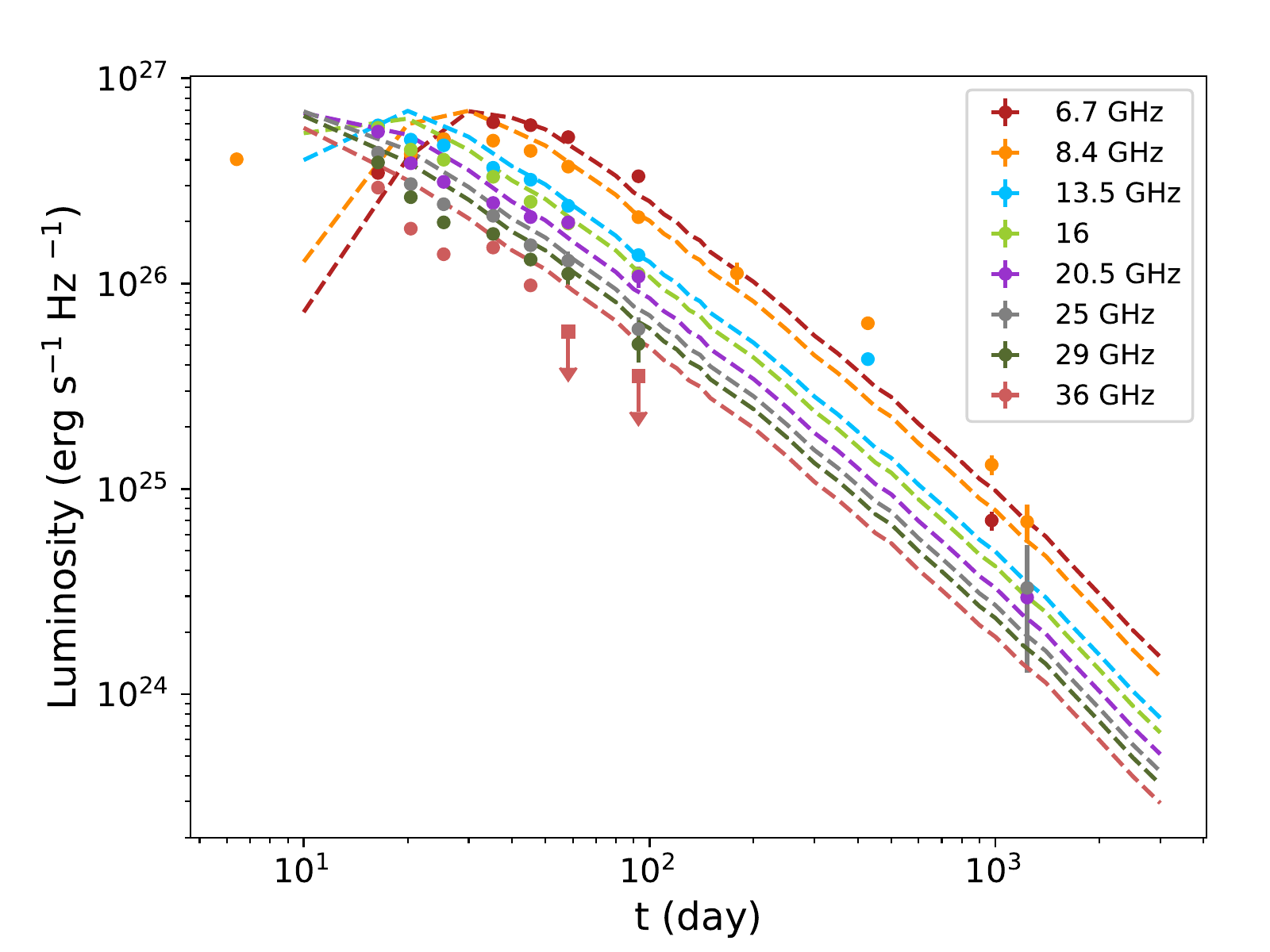}
\caption{Radio light curves of SN 2011dh in the range 300 MHz to 4.9 GHz (left panel) and between 6.7 and 36 GHz (right panel) for a wind-like ambient medium characterized by $\mdot = 4 \times 10^{-6}$ $\msunyr$ and $v_w =$ 10 $\kms$. The data are from \citet{sod12}, \citet{krauss12}, \citet{deWitt16} and this work.   
At high frequencies (right panel) our model has over-predicted the emission at early epochs. The reason for this overestimate is that we did not consider inverse Compton cooling in our model, as we are interested in studying the emission at late epoch. See text for details.}
%It is found from our study that the radio emission is consistent with a wind-like ambient medium, which is characterized by $\mdot/v_w = 4 \times 10^{-6}$ $\msunyr/ 10 \kms$. 
\label{fig:lightcurveSN11dh}
\end{figure*}

The light curves of SN~2011dh are shown in Figure~\ref{fig:lightcurveSN11dh}. The left panel shows the evolution of luminosity in the range 300 MHz to 4.9 GHz, while the right plot displays that between 6.7 and 36 GHz. It is found from our analysis that for this SN until $\sim$ 1300 days, up to the time archival data is available, since explosion the emission is consistent with a wind medium with a $\mdot/v_w = 4 \times 10^{-6}$ $\msunyr/ 10 \kms$. It is noted that our model predicts the low energy light curves (left panel of Fig.\ref{fig:lightcurveSN11dh}) with good accuracy, whereas in case of fluxes that are acquired at high frequencies (see the right panel of Fig.\ref{fig:lightcurveSN11dh}) the estimates are not that precise, especially at early epochs.
%It is noted that our model has better a fit for the low energy light curves (left panel of fig.\ref{fig:lightcurveSN11dh}) in comparison to the light curves acquired at high frequencies (see the right panel of fig.\ref{fig:lightcurveSN11dh}). 
The reason behind this discrepancy is that at high frequencies the emission becomes optically thin within a few days after the outbursts. At very early epoch, $\sim$ 20 days, inverse Compton (IC) loss plays a major role which steepen the electron spectrum and hence the radio flux decreases. In our radio calculation we have not considered the loss due to IC. Therefore, for high frequencies we have over-predicted the fluxes at early epochs. Furthermore, radiative cooling is not included in our hydro simulation. This is important at early time as described in $\S$\ref{sec:hydro_sim}, especially for SN~1993J. 
%There is another effect, the radiation cooling, which is not considered in our hydro simulation. This effect is also important at early time as described in $\S$\ref{sec:hydro_sim}.            

For SN~2011dh
%To fit the observed light curves 
we assumed $p = 2.95$ \citep{sod12, krauss12}, $\epsilon_{\rm e} = 0.03$ and $\epsilon_{\rm B} = 0.04$. At early times, if the radio emission does not come from a homogeneous spherical shell, the brightness temperature can not be represented by a fix temperature. It is found from our fitting that for SN 2011dh $T_{bright} = 3.9 \times 10^{9}~ {(t/1 ~ day)}^{0.32}$ K
%$T_{bright} = 3.9 \times 10^{9}~ t_{\rm day}^{0.32}$ K 
for $t \leq 1500$ days.
%, where $t_{\rm day}$ is the epoch since explosion. 
Beyond 1500 days the source will have a  brightness temperature of $4 \times 10^{10}$ K. It is noted that for this SN we do not need to take into account free-free absorption while calculating the radio fluxes, as the density of the CSM is low. The shock front radii and velocities predicted by our model at around 80 and 180 days after the explosion are found to be in good agreement with that estimated using the VLBI observations carried out at similar epochs \citep{bietenholz12}. 
%VLBI observations of this SN at 83 and 179 days after the explosion \citet{bietenholz12} estimated the shock front radii and shock velocities.   

\par
\subsection{X-rays}
\label{sec:Xray_result}
\subsubsection{SN 1993J}
\label{sec:Xray_93J}
\begin{figure}
\centering
\includegraphics[width=8.5cm,angle=0]{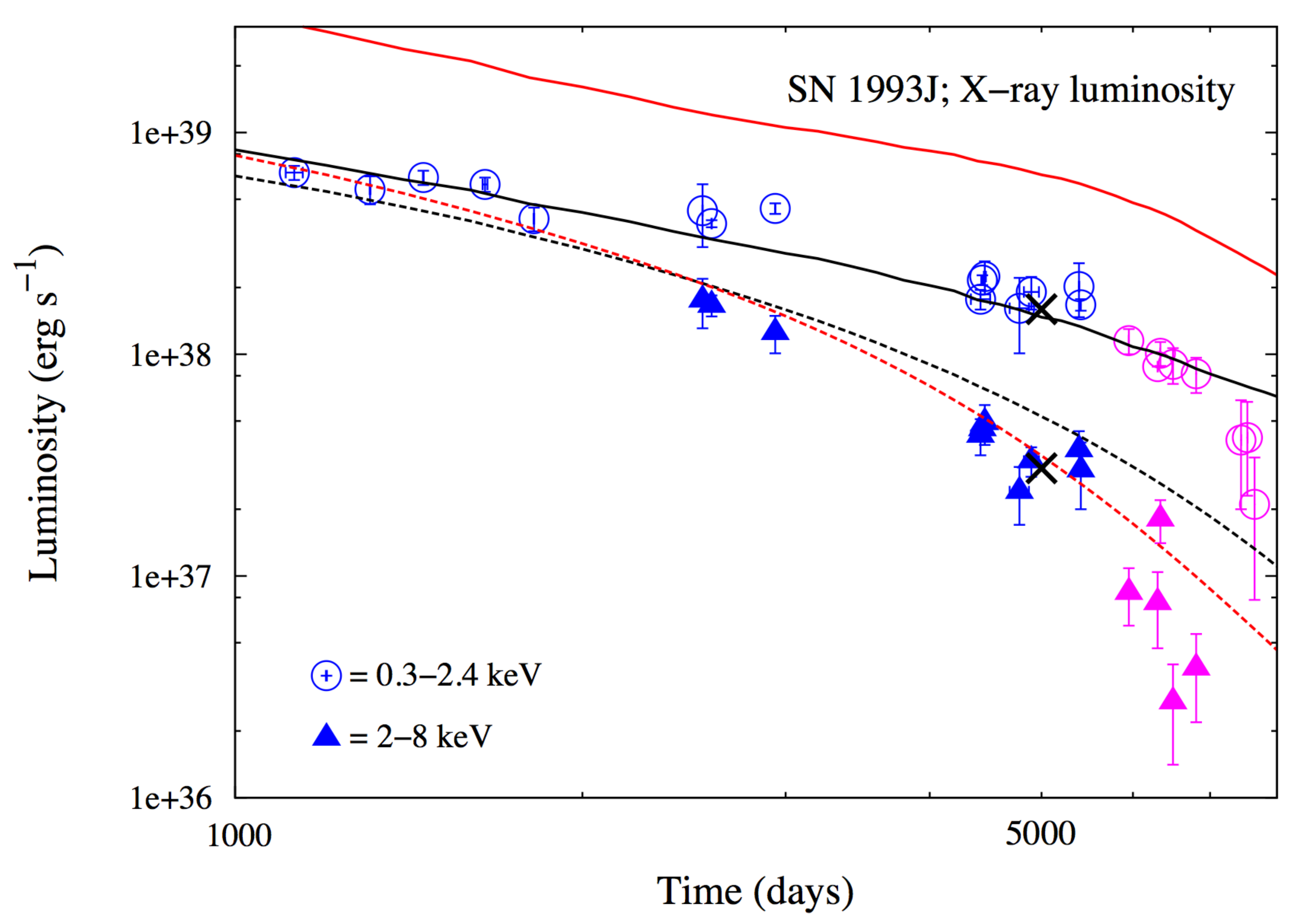}
\caption{Modeled X-ray light curves for SN~1993J. Data in blue are from \citet{chandra09}, and in magenta from \citet{dwa14}. Solid black line for the 0.3--2.4 keV emission from the reverse shock, and solid red line is for the same energy range, but assuming a factor of 3 extra compression of the shocked ejecta. Dashed black and red lines are the corresponding light curves for the 2--8 keV range. Note the relative insensitivity in emission due to this clumping effect for the 2--8 keV range. The two black crosses at 5000 days mark the effect of unequal temperatures of ions and electrons. These crosses relate to the black lines, but for the crosses we have put a cap on the electron temperature corresponding to 25\% of the shock temperature. Whereas the model can adequately fit the light curves up to $\sim 7000$ days, the apparent faster downturn thereafter is not modeled well. A likely explanation is a flatter ejecta structure encountered by the reverse shock than in the model simulations (see text).}
\label{fig:Xray_93J}
\end{figure}

For SN 1993J, X-ray data until 5408 days are compiled in \citet{chandra09}. Additional data up to day $\sim$ 7650 are plotted in \citet{dwa14}. All data are shown in Figure~ \ref{fig:Xray_93J} together with our model results using the ejecta model described in $\S$ \ref{sec:ejecta_model}, and a wind density characterized
by $4\EE{-5} (v_{\rm w}/10~{\kms})~{\rm M}_\odot~{\rm yr}^{-1}$, i.e., ten times higher circumstellar density than for our SN~2011dh simulations. 
%The mass loss rate used for SN~1993J is similar to what was used by \citet{fra96}. 
Beyond a radius of $\sim 2 \times 10^{17}$ cm, we assume a steeper density profile in the circumstellar medium than $\rho_{w} \propto r^{-2}$. A steeper density profile was also discussed by \citet{chandra09} who argue for $\rho_{w} \propto r^{-2.6}$ for the last epochs discussed in their paper.

\par 
%We use the same type of modeling for the X-ray emission of SN~1993J as was discussed for SN~2011dh in \S 7.2.1. 

Since we do not include radiative cooling in the hydrodynamic simulations, we have concentrated on epochs later than 1000 days (see also below). We show two cases, where the black line depicts the emission from the plasma shocked by the reverse shock and the red one exhibits the effect of clumping. For clumping we assume that all the gas shocked by the reverse shock is 
compressed further by a factor of three (and has a factor of three lower filling factor). It is considered in our model that the gas is in pressure equilibrium so that the temperature of the compressed plasma is a factor of three lower than for the original model. 
As can be seen in Figure~\ref{fig:Xray_93J}, our model simulates the emission fine for the light curve of soft X-ray emission up to $\sim 7000$ days, without invoking any type of clumping of the shocked ejecta (solid black line). Including an extra compression by a factor of 3 increases the soft X-ray emission significantly. %%(solid red line). 
Although clumping is, from many perspectives, much more complicated than just applying a compression factor, this gives a hint on how the emission can be affected by clumping (also see Figure~\ref{fig:Xray_11dh} which demonstrates the impact of clumping on the X-ray spectrum).  
For the hard X-ray emission, the difference between the two models (i.e., no extra compression, and an extra factor-of-three compression) is small.  Variations in the soft X-ray light curve between $1000-3000$ days could therefore be due to modest density inhomogeneities at the reverse shock, rather than invoking variations of the circumstellar density with radius, as was suggested by \citet{chandra09}. A circumstellar density drop coinciding in radius with $r_{\rm s} \sim 2 \EE{17}~\rm{cm}$ may be needed, but the shallower ejecta profile encountered by the reverse shock also plays a role.

\begin{figure}
\centering
\includegraphics[width=8.7cm,angle=0]{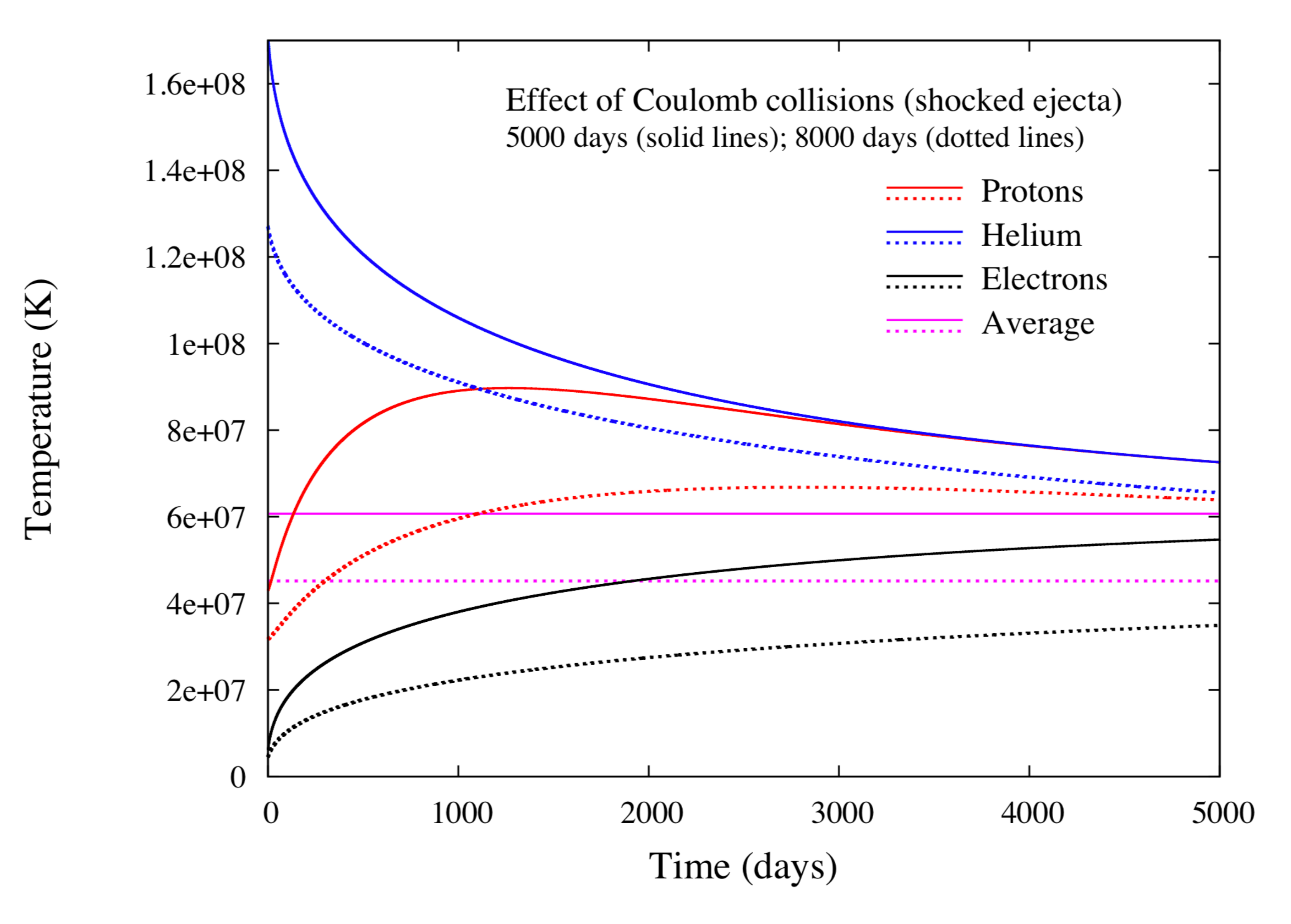}
\caption{Temperatures for a fully ionized H-He plasma, simulating the shocked ejecta of SN~1993J at 5000 days (solid lines) and 8000 days (dotted lines). Blue is for helium, red is for protons and black is for electrons. Only Coulomb collisions are considered, except for electrons which are assumed to attain 10\% of the average shock temperature immediately behind the shock front. Density and abundances are assumed to be constant with time since the gas was shocked. At these epochs, the He/H-ratio (by number) in our simulation of SN~1993J is $\sim 38$ and $\sim 94$, respectively. Note the long time for temperature equilibration of electrons and ions.}
\label{fig:Equil_93J}
\end{figure}

If $t_{\rm cool}$ and $t_{\rm dyn}$ represent the cooling and dynamic time scales, respectively, then our assumption of an adiabatic reverse shock for SN~1993J is justified by $t_{\rm cool}/t_{\rm dyn} > 1$ of newly shocked ejecta at all our epochs ($t > 1000$~days), even if an extra compression factor of 3 is added. The same is actually true for the full shock structure all the way out to the contact discontinuity if we do not add any extra compression. For the extra compression of 3, $t_{\rm cool}/t_{\rm dyn} < 1$ for the shocked ejecta closest to the contact discontinuity for all epochs considered (i.e., up to 8000 days). Since Figure~\ref{fig:Xray_93J} indicates that no significant clumping is needed, we believe that an adiabatic shock structure is adequate. While the soft X-ray luminosity agrees reasonably with observations out to $\sim 7000$ days, the emission from hard X-rays is overproduced in our model. This is most likely not an effect of clumping, but could be due to unequal electron and ion temperatures ($T_{\rm e}$ and $T_{\rm i}$, respectively). For example, at 5000 days, our simulations give a reverse shock temperature which is in the range  $(0.7-6.1)\EE{7}$~K. Limiting $T_{\rm e}$ so that it instead is in the range $(0.7-1.5)\EE{7}$~K, decreases the hard X-ray flux by a factor of $\sim 1.7$, while the soft X-ray flux is increased by $\sim$ 3\%. This is shown with black crosses in Figure~ \ref{fig:Xray_93J}, and argues for $T_{\rm e} / T_{\rm i} \sim 0.25$ at the reverse shock, which is in line with the assumption of  $T_{\rm e} / T_{\rm i} \sim 0.15$ by \citet{chandra09} to produce hard X-ray emission at adequate levels. For SN~1993J at 5000 days, in our model, the reverse shock is driven into helium-dominated ejecta, and the equipartition time scale due to Coulomb collisions is $t_{\rm eq} \sim 3.4\times 10^3~(T_{\rm e}/10^7 {\rm K})^{3/2}~(n_{\rm He}/10^3 \cm3)^{-1}$~days \citep{spit62}. In our model, the helium density is $\sim 4.1\EE3 \cm3$ ($1.2\EE3 \cm3$) at 5000 (8000) days, so that $t_{\rm eq} <  t_{\rm dyn}$. This is shown in Figure~\ref{fig:Equil_93J}. While Coulomb collisions are not able to fully equilibrate ion and electron temperatures at the shock front, a ratio of $T_{\rm e} / T_{\rm i} = 0.15 $ seems to be on the low side, unless the ejecta density is lower than in our model. This could be the case if the slope of the ejecta profile is flatter than in our model at these epochs. The positions of the reverse shock in our ejecta model at 1000, 2000, 3000, 5000, 7000 and 8000 days are shown with dotted lines in the left panel of Figure~ \ref{fig:ejectastruc}.

\subsubsection{SN 2011dh}
\label{sec:Xray_11dh}
\begin{figure}
\centering
\includegraphics[width=8.7cm,angle=0]{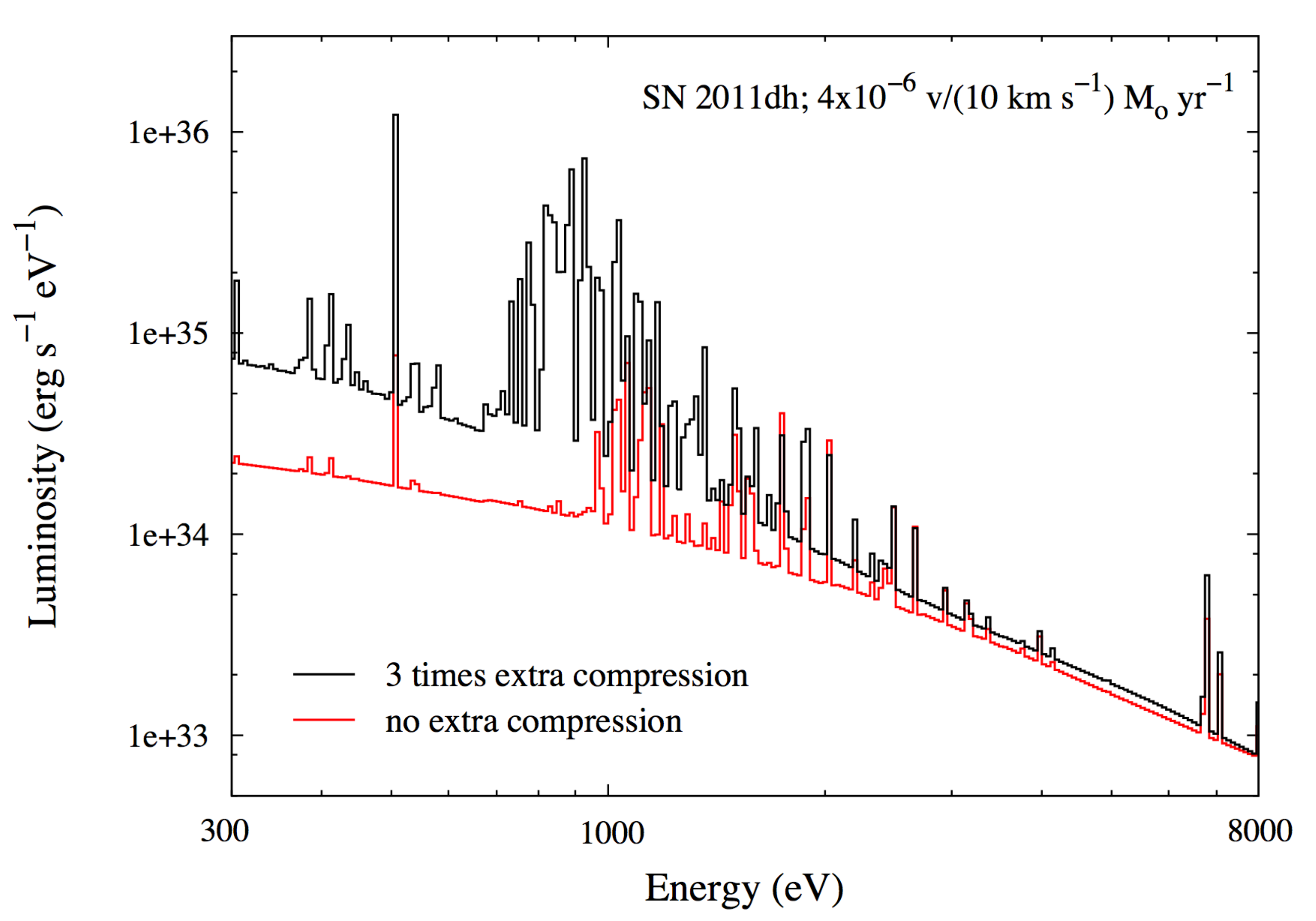}
\caption{Modeled X-ray emission from SN~2011dh at 500 days after explosion. Spectrum in red is from the reverse shock using parameters from our hydro models. The spectrum in black is for an assumed extra compression of the shocked ejecta by a factor of three, considering pressure equilibrium. This gives an indication of the effect of clumping in the shocked ejecta. Note that the spectra above $\sim 2$ keV are very similar.}
\label{fig:Xray_11dh}
\end{figure}

For SN 2011dh we show in Figure~\ref{fig:Xray_11dh} our modeled X-ray emission at 500 days for a wind characterised by a mass loss rate of $4\EE{-6} (v_{\rm w}/10~{\kms})~{\rm M}_\odot~{\rm yr}^{-1}$. 
As for SN 1993J, we consider two cases, where the spectrum in red exhibits the emission from the shocked ejecta without assuming a compression factor. The spectrum is featureless and dominated by free-free emission, as expected given the high temperature of the shocked ejecta, $\sim 3.5\times10^8$~K. The black spectrum in Figure~\ref{fig:Xray_11dh} shows the effect of clumping. This figure demonstrates that clumping influences the spectrum at energies below $\sim 2$ keV. That is why, in case of SN 1993J, we see small differences in hard X-ray luminosity (2-8 keV) calculated without considering any compression and invoking a clumping of a factor of 3 (black and red dashed lines in fig. \ref{fig:Xray_93J}).    
%For this we have assumed that the all the gas shocked by the reverse shock is compressed further by a factor of three (and has a factor of three lower filling factor). We assume that the gas is in pressure equilibrium so that the temperature of the compressed plasma is a factor of three lower than for the original model (red spectrum). Although clumping is, from many perspectives, much more complicated than just applying a compression factor, the models in Figure~\ref{fig:Xray_11dh} give a hint on how the spectrum could be affected by clumping. In particular, the models in this figure show that clumping is expected to show large differences in the spectrum, especially at energies below $\sim 2$ keV. 

\par 
Clumping decreases the cooling time, $t_{\rm cool}$, of the shocked gas, making it more likely to become shorter than the dynamical time scale, $t_{\rm dyn}$. Looking more into detail of the two compression factors (i.e., 1 and 3) in the models in Figure~ \ref{fig:Xray_11dh}, $t_{\rm cool}$ of the just shocked ejecta is $\sim 9.9\times10^5$ days and $\sim 1.9\times10^5$ days, respectively. Since $t_{\rm dyn}$ is 500 days, this means that shocks driven into the ejecta are not radiative, which justifies our model assumption of an adiabatic shock. This does not necessarily mean that radiative cooling could not affect the shock structure. Close to the contact discontinuity, the cooling times are $\sim 1.5\times10^3$ days and $\sim 1.3\times10^2$ days, respectively, which means that for the factor-of-3-compressed model, radiative cooling close to the contact discontinuity, where the temperature gets below $\sim 10^7$~K, could marginally affect the modeled spectrum. 

\par 
For the two clump compression factors, 1 and 3, the post-processed spectral synthesis using our adiabatic models gives luminosities from the reverse shock that are $\sim 6.3 \times 10^{37}$ erg~s$^{-1}$ and $\sim 2.1 \times 10^{38}$ erg~s$^{-1}$, respectively. This is much less than for a fully radiative reverse 
shock, $L_{\rm rev} = 2 \pi r_{\rm ej}^2 \rho_{\rm ej} v_{\rm rev}^3$, which for our SN~2011dh model at 500 days is  $\sim 1.0 \times 10^{40}$ erg~s$^{-1}$. Here $r_{\rm ej}$ and $\rho_{\rm ej}$ represent the radius and density of the outer most unshocked ejecta, respectively. $v_{\rm rev}$ is the reverse shock velocity. This estimate indeed shows that radiative cooling is not important for the reverse shock of SN~2011dh at 500 days. In the range $0.3-8.0$ keV the luminosities are 
$\sim 4.1 \times 10^{37}$ erg~s$^{-1}$ and $\sim 1.5 \times 10^{38}$ erg~s$^{-1}$ for the compression factors 1 and 3, respectively. The observed luminosity in the range $0.3-8.0$ keV is $\sim 6 \times 10^{37}$ erg~s$^{-1}$ \citep{maeda14}, which indicates that only modest clumping of the gas being shocked is needed for our model with $4\EE{-6} (v_{\rm w}/10~{\kms})~{\rm M}_\odot~{\rm yr}^{-1}$ to reproduce the observed emission. The analysis by \citet{maeda14} suggested a wind density roughly half that in our analysis. The difference stems from an assumed high power-law index, $n = 20$, for the density of the supernova ejecta, $\rho_{\rm ej} \propto r^{-n}$, in \citet{maeda14}. In the self-similar solution \citep{che82a}, $\rho_{\rm ej}/\rho_{\rm csm}$ is a strong function of $n$. 
%(Here $\rho_{\rm w}$ is the preshock circumstellar density.) 
In our model, $n \approx 7$ around 500 days, but the structure of the shocked ejecta has a memory of a much steeper profile of the outermost ejecta being shocked at earlier times (cf. right panel of Figure~ \ref{fig:ejectastruc}). The position of the reverse shock in our ejecta model at 500 days, along with that at epochs 200 and 1200, is shown with dotted line in the right panel of Figure~ \ref{fig:ejectastruc}.  

%\begin{figure}
%\centering
%\includegraphics[width=8cm,angle=0]{93J_Xrayfigure.pdf}
%\caption{\bf Please change the x-axis caption. It should not be log(luminosity). }
%\label{fig:Xray_93J}
%\end{figure}

%The above equation (eq. \ref{eq:csm2})   

 %Therefore, beyond $r = 2 \times 10^{17}$ cm the density star   

\section{Discussion: Binary evolution of progenitors}
\label{sec:discussion}

The progenitors of SNe 1993J and 2011dh were very likely the part of a binary system. Although the binary companion of SN 1993J has not yet been discovered exclusively, the studies carried out during the last 25 years \citep{cohen95, nomoto93,pod93,aldering94,maund04,maund09,fox14} are pointing toward the presence of a massive hot companion star in the vicinity of the SN. The proposed model for the SN 1993J progenitor system consists of two stars of comparable masses, $\sim$ 15 $\msun$ each \citep{nomoto93, pod93, maund04}, and in which the progenitor star loses mass to the binary companion though the stable case C mass transfer \citep{pod92}. After the exhaustion of the core helium, when the progenitor is on the asymptotic giant branch, the primary fills its Roche lobe and starts to transfer mass to the companion. The secondary accretes most of the transferred material and the remaining fraction gets lost in the surrounding medium. The mass transfer ceases when the primary has transferred nearly all of its hydrogen envelope. At this  point the hydrogen mass in the envelope of the primary becomes less than a certain level and the star starts to shrink. As a consequence, it is now not possible for the primary to fill its Roche lobe. The primary then slowly loses the residual hydrogen envelope through stellar winds. For a $\sim 0.5 \msun$ residual envelope mass, and a mass-loss rate of $4 \times 10^{-5}$ $\msunyr$, it will take around  $12 \times 10^{3}$ years to completely strip the envelope. This time is enough for the evolved star to go off as SN. 
%Therefore, when the star exploded, a tiny amount of envelope hydrogen was left. Moreover, the companion star now has become a very luminous star of mass of $\sim$ 22 $\msun$ after accreting a considerable amount of mass from the progenitor star.    

%If this the case then the immediate surrounding of the SN has been shaped by the wind from progenitor star. 

From our radio and X-ray studies we have indications that around 6500 years  before the explosion of  SN 1993J, the density of the surrounding medium starts to increase rapidly towards the time leading up to the explosion. This implies that before $\sim 10^{4}$ years prior the explosion, the mass loss mechanism was most likely different from that due to the wind from the primary star. It is possible that before this time the primary mainly loses mass to the secondary companion through Roche lobe overflow, and because of the high accretion efficiency of the companion, a very small percentage of the accreted material was managed to escape from the system. For a star of main sequence mass of $\sim 15$ $\msun$ the central carbon burning phase starts around $\sim 10^{4}$ years prior the explosion. Therefore, it is possible that the primary ceases the mass transfer through Roche lobe overflow from the time the carbon at the core starts to fuse. However, according to the \citet{woosley94} at the time of the explosion, the primary was big enough so that it was able to transfer mass to the secondary star through Roche lobe overflow.  If this is the case, then it suggests that at the onset of the core carbon burning phase due to some reason the accretion efficiency of the companion star decreases. As a consequence, more mass gets lost in to the circumbinary medium and a comparatively dense ambient medium forms. It is, however, difficult to explain what exactly causes the accretion efficiency to change so abruptly around $t \sim 10^{4}$ years prior to the explosion.  

There exists another possibility. \citet{maund04} found that the binary companion of SN 1993J was very likely to be a B2 Ia star. For this kind of star the mass loss rate is in the range (0.25 - 1) $\times 10^{-6}$ $\msunyr$ for a wind velocity of $\sim 600 \kms$ \citep{kudritzki99,crowther08}. In our model we assumed that the density of the surrounding medium for t $<< 10^{4}$ years (i.e. $r > 2 \times 10^{18}$ cm ($\equiv R_{\rm out}$)) can be characterized by $\mdot = 1.2 \times 10^{-5} $ $\msunyr$ for a wind velocity of $600 \kms$ (i.e., $\mdot = 2 \times 10^{-7} $ $\msunyr$ for $v_w = $ 10 km/sec), which is around 25 times higher than that expect from a B2 Ia star. However, the derived $\mdot$ and $v_w$ for these stars are not independent of theoretical assumptions. It is found from our study that around 25 years after the explosion the radius of the SN is $\sim 9.4 \times 10^{17}$ cm. To probe a region beyond $R_{\rm out}$ the SN will take another $\sim$ 25 years. It is therefore possible that the density around $R_{\rm out}$ is lower than what we have presumed in our model, and it has been due to the winds from the companion star. Thus the CSM before the carbon burning phase of the primary could have been created by the winds from the companion star, and that from carbon burning until the explosion had been due to the mass loss through Roche lobe which gave rise to a comparatively dense surroundings.

\par
In case of SN 2011dh, a yellow supergiant star was observed at the SN location in the pre-explosion HST archival imaging \citep{van11}. However, it was initially thought that the yellow supergiant was not the star that exploded rather a neighbouring or companion of the progenitor \citep{van11, arcavi11, sod12}. Later, \citet{van13} conducted a search for the companion star of SN 2011dh $\sim$ 650 days after the explosion using the HST WFC3, and found that the yellow supergiant had disappeared from the SN location. This implied that it was the supergiant that exploded and resulted in a type IIb SN. From optical and near infrared photometry and spectroscopy of the SN, at early and late epochs, and modeling of the lightcurves \citet{ergon14,ergon15} suggested that SN 2011dh was part of a binary system. It is interesting that analysing HST data \citet{maund15,folatelli14} have detected a point like object, at all wavelengths, at the location of the SN.
%From optical and near infrared photometry and spectroscopy of the SN, at early and late epochs, and modeling of the lightcurves \citet{ergon14,ergon15} estimated a helium core mass of 3-4 $\msun$ and an upper bound on the MS mass of $\sim$ 15 $\msun$ for SN 2011dh. These authors therefore suggested that SN 2011dh was part of a binary system. According to \citet{ben13} if the mass transfer efficiency of the primary is close to 0.25, the putative companion should have been possible to detect in the UV $\sim$ 900 days after the explosion. Nevertheless, this estimate critically depends on the the efficiency at which mass had been transferred to the secondary before explosion. Therefore, in case the efficiency is low, one is required to wait several years before getting a confirmation signature from the possible companion. It is interesting that analysing HST data \citet{maund15,folatelli14} have detected a point like object, at all wavelengths, at the location of the SN.

\par 
From our study we found that the radio and the X-ray emission from SN~2011dh is consistent with a wind medium (see Fig. \ref{fig:lightcurveSN11dh}) characterised by a mass-loss rate of $ 4 \times 10^{-6}$ $\msunyr$ for a wind velocity of $10 \kms$. This value of $\mdot/v_w$ is almost 10 times higher than that predicted by \citet{sod12,krauss12} by analyzing the early radio and X-ray light curves. However, to explain the thermal X-ray emission at around $\sim 500$ days post explosion \citep{maeda14} was required to adopt a high density medium with a $\mdot \sim 3 \times 10^{-6} \msunyr (v_w/20 \kms)^{-1}$ around the progenitor. As mentioned in \S \ref{sec:Xray_11dh} this estimate depends on the density slope of the ejecta model used, and we argued that our results, and those of \citet{maeda14} are consistent. A likely reason for the difference between us and \citet{sod12,krauss12} is the high values of $\epsilon_{\rm B}$ and $\epsilon_{\rm e}$ used in the studies of those authors.

A star usually spends around $\sim 3000$ years in the yellow supergiant phase \citep{drout09} before exploding, and the mass-loss rate at this phase is highly unconstrained. \citet{georgy11} carried out calculations of a single $\sim$ 14 $\msun$ rotating stellar model and assumed a 3-10 times higher mass loss from the stars at the red supergiant (RSG) phase. It was demonstrated in this study that with a higher value of $\mdot$ the stars end their nuclear life at a position in the Hertzsprung-Russell diagram which is roughly consistent with the observed position of several yellow supergiant progenitors. Therefore, it may be that SN~2011dh was an explosion of a yellow supergiant which was not a part of a binary system and had undergone mass loss through wind at a high rate during its red supergiant phase so that it had lost almost the entire hydrogen envelope before explosion. It is at present not ruled out that a YSG in solitude could be responsible for SN~2011dh. However, it is difficult from our analysis to distinguish between the single and binary progenitor scenario. With more observations in the near future, one may be able to disentangle this issue.    

For the radio emission of both SNe~1993J and 2011dh, VLBI radio imaging can give clues to the details of the circumstellar interaction. The details are best explored for SN~1993J, and radio images up to $\sim 10$ years show that both the inner and outer edges of the radio shell expand roughly as $\propto t^{m}$, with $m\simeq 0.85-0.87$ \citep{marca09,Mart11b}. We find $m\sim 0.85$ for this period, in agreement with the estimates by \citet{marca09} and \citet{Mart11b}, while for epochs after $\sim 2500$ days, the reverse shock sligthly lags behind this expansion and the forward shock speeds up. The former is due to the ejecta density slope being shallower at the position of the reverse shock, and the latter is because of the sharp density decrease in the CSM. 
%The observed lag of the reverse shock after $\sim 10$ years compared to $\propto t^{0.85}$ is more pronounced than in our models, which suggests that the ejecta density slope could be flatter than in the explosion model we have used. 
At $\sim 5000$ days, the reverse shock encounters ejecta with $n \sim 2.5$.
%, and for the SN~1993J this is most likely too steep.
We will explore this and its ramifications for radio and X-ray emission in greater detail in future analyses. A discussion on how the radio emission may depend on the effects of a weakening reverse shock is discussed by \citet{bjornsson15}.

\section{Conclusions}
\label{sec:conclusions}
We have performed hydrodynamical simulations of the interaction between SN ejecta and the CSM. The SNe we have modeled are SNe~1993J and 2011dh. For our study we have taken the ejecta structures of these two SNe from numerical simulations (STELLA). The main aim of our study is to try to trace the mass loss history of the progenitors, and hence gain information about the evolution of the progenitor systems. SN shocks are often bright in radio and X-ray wavelengths. 
While the radio emission has been assumed to mainly originate at the region close to the forward shock, the X-rays at late epochs predominantly come from the shocked ejecta behind the reverse shock. There is, however, considerable uncertainty regarding the exact location of the radio emission, and VLBI imaging of SN 1993J actually shows a rather even distribution of radio emission for the full region between the two shocks. We have adopted this approach, i.e., we assume homogeneous radio emission from the shocked CSM. For the X-rays, we assume that the sole produced is the reverse shock. Although radio and X-ray emission emanate from different parts of the shocked gas, the flux of both depends on the density and the structure of the surrounding medium, as well as the slope of the density profile of the SN ejecta. How dense the surrounding medium is depends on the amount of mass ejected by the progenitor system during its evolution. Therefore, with accurate information about the evolution of SN shocks, a reliable ejecta model, and proper modeling of radio and X-ray emissions one can map the mass-loss rates from the progenitor system at different phases of its evolution.

\par 
%Here we have presented a combined study of late time radio and X-ray emissions from SN 1993J and SN 2011dh. 
The late time radio and X-ray curves of SN~1993J have shown a sudden downturn in radio and X-ray fluxes beyond $\sim$ 3000 days after the explosion of the SN.
%as well as an 
Also the inner edge of the radio shell starts to lag compared to the $\propto t^{0.85}$ evolution at earlier epochs. Evaluating the SN evolution through hydro simulation, and studying late time radio and X-ray emissions from SN~1993J, we found that to account for the observed drops, the density of the CSM needs to decrease rapidly at $r > 2 \times 10^{17}$ cm. For smaller radii the wind density is characterized by $\mdot = 4 \times 10^{-5}$ $\msunyr$ for $v_w$ = 10 $\kms$. This implies that if the primary transfers mass to the companion, through Roche lobe overflow from the end of the core helium burning stage until the explosion, then $\sim 6500~(v_w / 10 \kms)^{-1}$ years before the explosion the accretion efficiency of the secondary decreases. Therefore, more mass gets ejected in the ambient medium which makes it denser. For a primary, with a $\sim$ 15 $\msun$ main sequence mass, it is expected that around $10^4$ yrs prior to the explosion the star was burning the carbon present at its core. However, to figure out the reason behind this decrease in accretion efficiency of the companion star is beyond the scope of this paper. %From the apparent lag of the reverse shock compared to a simple power law, it seems unavoidable that the density slope of the SN ejecta is shallower than in the ejecta model we have used. 

\par 
An important future investigation is also to test the scenario discussed by \citet{bjornsson15}. He suggests that the seemingly achromatic break in radio and X-ray light curves around $\sim 3000$ days could be due to the reverse shock entering a flat part of the ejecta density profile. When this happens, the reverse shock would weaken, or perhaps even disappear. This should reduce the X-ray emission and cause a decline in the radio emission. In our model, the density profile of the ejecta just about to be shocked around 3000 days is $n=3.1$. In the scenario envisioned by Bj\"ornsson, the density profile is even shallower. This would produce weaker
X-ray emission than in our model, and could be a reason why our X-ray light curves in Figure \ref{fig:Xray_93J} overshoot at $t > 5000$ days.

\par
In case of SN~2011dh both radio and X-ray emissions are consistent with a wind-like ambient medium. From HST WFC3 imaging around 650 days after the explosion, \citet{van13} confirm that the YSG discovered in the pre-explosion archival HST image was the progenitor of the SN. With the SN ejecta structure from STELLA we found that to account for both the radio and the X-ray fluxes at late time we require $\mdot = 4 \times 10^{-6}$ $\msunyr$ for $v_w$ = 10 $\kms$. It is, however, difficult from our study to conclude whether the YSG had evolved in solitude, or was a part of a binary system. Nevertheless, in future, if we observe a drop in radio and X-ray emission from SN 2011dh, similar to that has been observed from SN~1993J, then that would suggest a scenario where the progenitors of both SNe~1993J and 2011dh had undergone similar kind of binary evolution before explosion.                    

\section{Acknowledgement}
We thank Claes-Ingvar Bj{\"o}rnsson, Michiel Bustraan and Claes Fransson for discussions. The software used in this work was in part developed by the DOE NNSA-ASC OASCR Flash Center at the University of Chicago. The hydrodynamical simulations shown were performed on resources provided by the Swedish National Infrastructure for Computing (SNIC) at PDC (Beskow), Royal Institute of Technology, Stockholm. PC acknowledges support from the Department of Science and Technology via SwaranaJayanti Fellowship award (file no. DST/SJF/PSA-01/2014-15).  
%Before 6500 years befor the explosion the dens             

%While the flux of the radio emission is roughly proportional to the density of the ambient medium, the X-ray emission at late epoch, predominantly comes from shocked ejecta behind the reverse shock, gets affected by the nature of the CSM               

\end{document}